\documentstyle{mn}

\newif\ifAMStwofonts

\ifoldfss
  \ifCUPmtlplainloaded \else
    \NewTextAlphabet{textbfit} {cmbxti10} {}
    \NewTextAlphabet{textbfss} {cmssbx10} {}
    \NewMathAlphabet{mathbfit} {cmbxti10} {} 
    \NewMathAlphabet{mathbfss} {cmssbx10} {} 
  \fi
  \ifAMStwofonts
    \ifCUPmtlplainloaded \else
      \NewSymbolFont{upmath} {eurm10}
      \NewSymbolFont{AMSa} {msam10}
      \NewMathSymbol{\upi}     {0}{upmath}{19}
      \NewMathSymbol{\umu}     {0}{upmath}{16}
      \NewMathSymbol{\upartial}{0}{upmath}{40}
      \NewMathSymbol{\leqslant}{3}{AMSa}{36}
      \NewMathSymbol{\geqslant}{3}{AMSa}{3E}

    \fi
  \fi
\fi 

\ifnfssone
  \newmathalphabet{\mathit}
  \addtoversion{normal}{\mathit}{cmr}{m}{it}
  \addtoversion{bold}{\mathit}{cmr}{bx}{it}
  \newmathalphabet{\mathbfit} 
  \addtoversion{normal}{\mathbfit}{cmr}{bx}{it}
  \addtoversion{bold}{\mathbfit}{cmr}{bx}{it}
  \newmathalphabet{\mathbfss} 
  \addtoversion{normal}{\mathbfss}{cmss}{bx}{n}
  \addtoversion{bold}{\mathbfss}{cmss}{bx}{n}
  \ifAMStwofonts
    \ifCUPmtlplainloaded \else
      \UseAMStwoboldmath
      \makeatletter
      \new@mathgroup\upmath@group
      \define@mathgroup\mv@normal\upmath@group{eur}{m}{n}
      \define@mathgroup\mv@bold\upmath@group{eur}{b}{n}
      \edef\UPM{\hexnumber\upmath@group}
      \new@mathgroup\amsa@group
      \define@mathgroup\mv@normal\amsa@group{msa}{m}{n}
      \define@mathgroup\mv@bold\amsa@group{msa}{m}{n}
      \edef\AMSa{\hexnumber\amsa@group}
      \makeatother
      \mathchardef\upi="0\UPM19
      \mathchardef\umu="0\UPM16
      \mathchardef\upartial="0\UPM40
      \mathchardef\leqslant="3\AMSa36
      \mathchardef\geqslant="3\AMSa3E
    \fi
  \fi
\fi 

\ifnfsstwo
  \DeclareMathAlphabet{\mathbfit}{OT1}{cmr}{bx}{it}
  \SetMathAlphabet\mathbfit{bold}{OT1}{cmr}{bx}{it}
  \DeclareMathAlphabet{\mathbfss}{OT1}{cmss}{bx}{n}
  \SetMathAlphabet\mathbfss{bold}{OT1}{cmss}{bx}{n}
  \ifAMStwofonts
    \ifCUPmtlplainloaded \else
      \DeclareSymbolFont{UPM}{U}{eur}{m}{n}
      \SetSymbolFont{UPM}{bold}{U}{eur}{b}{n}
      \DeclareSymbolFont{AMSa}{U}{msa}{m}{n}
      \DeclareMathSymbol{\upi}{0}{UPM}{"19}
      \DeclareMathSymbol{\umu}{0}{UPM}{"16}
      \DeclareMathSymbol{\upartial}{0}{UPM}{"40}
      \DeclareMathSymbol{\leqslant}{3}{AMSa}{"36}
      \DeclareMathSymbol{\geqslant}{3}{AMSa}{"3E}
    \fi
  \fi
\fi 

\ifCUPmtlplainloaded \else
  \ifAMStwofonts \else 
    \def\upi{\pi}
    \def\umu{\mu}
    \def\upartial{\partial}
  \fi
\fi

\title[Extragalactic radio sources at 20\,GHz] {The properties 
of extragalactic radio sources selected at 20\,GHz }

\author[Sadler et al. ]{
\parbox[t]{\textwidth}{
Elaine M.\ Sadler$^1$, 
Roberto Ricci$^2$, 
Ronald D.\ Ekers$^2$, 
J.A.\ Ekers$^2$, 
Paul J.\ Hancock$^1$, 
Carole A.\ Jackson$^2$,
Michael J.\ Kesteven$^2$, 
Tara Murphy$^1$, 
Chris Phillips$^2$\,
Robert F.\ Reinfrank$^3$, 
Lister Staveley--Smith$^2$, 
Ravi Subrahmanyan$^2$, 
Mark A.\ Walker$^{1,2,4}$, 
Warwick E.\ Wilson$^2$,
Gianfranco De Zotti$^{5,6}$ 
}
\vspace*{6pt} \\
$^1$School of Physics, University of Sydney, NSW 2006, Australia\\
$^2$Australia Telescope National Facility, CSIRO, P.O. Box 76, Epping,
    NSW 1710, Australia \\
$^3$Department of Physics and Mathematical Physics, University of Adelaide, 
Adelaide, SA 5005, Australia \\
$^4$ MAW Technology Pty Led, 3/22 Cliff St., Manly 2095, Australia \\
$^5$SISSA/ISAS, Via Beirut 2--4, I-34014 Trieste, Italy \\
$^6$INAF, Osservatorio Astronomico di Padova, Vicolo
dell'Osservatorio 5,
I-35122 Padova, Italy \\
}

\pagerange{\pageref{firstpage}--\pageref{lastpage}}
\pubyear{2000}

\def\nodata{ ~$\cdots$~ }

\begin{document}

\maketitle

\label{firstpage}

\begin{abstract}
We present some first results on the variability, polarization and general 
properties of radio sources selected at 20\,GHz, 
the highest frequency at which a sensitive radio survey has been carried 
out over a large area of sky. 
Sources with flux densities above 100\,mJy in the ATCA 20\,GHz Pilot 
Survey at declination $-60^\circ$ to $-70^\circ$ were observed at 
up to three epochs during 2002--4, including near-simultaneous measurements 
at 5, 8 and 18\,GHz in 2003.  
Of the 173 sources detected, 65\% are candidate QSOs or 
BL Lac objects, 20\% galaxies and 15\% faint 
($b_{\rm J}>22$\,mag) optical objects or blank fields. 

On a 1--2\,year timescale, the general level of variability at 
20\,GHz appears to be low.  
For the 108 sources with good--quality measurements in both 2003 and 2004, 
the median variability index at 20\,GHz was 
6.9\% and only five sources varied by more than 30\% in flux density. 

Most sources in our sample show low levels of linear 
polarization (typically 1--5\%), with a median fractional polarization 
of 2.3\% at 20\,GHz.  There is a trend for fainter 20\,GHz sources 
to have higher fractional polarization.

At least 40\% of sources selected at 20\,GHz have strong spectral 
curvature over the frequency range 1--20\,GHz.  
We use a radio `two--colour diagram' to characterize the radio 
spectra of our sample, and confirm that the flux densities of 
radio sources at 20\,GHz (which are also the foreground point-source 
population for CMB anisotropy experiments like WMAP and Planck) cannot 
be reliably predicted by extrapolating from surveys at lower frequencies.  
As a result, direct selection at 
20\,GHz appears to be a more efficient way of identifying 90\,GHz 
phase calibrators for ALMA than the currently--proposed technique of 
extrapolation from radio surveys at 1--5\,GHz. 
\end{abstract}

\begin{keywords}
surveys --- cosmic microwave background --- 
galaxies: radio continuum --- galaxies: active
--- radio continuum: general
\end{keywords}


\section{Introduction}
Most large--area radio imaging surveys have been carried out at 
frequencies of 1.4\,GHz or below, where the long--term variability 
of the radio--source population is generally low. 
The catalogued flux densities measured by such surveys can 
therefore continue to be used with a high level of confidence 
for many years after the survey was made. 

It is not clear to what extent this is true for radio surveys 
carried out at higher frequencies, where the source population 
becomes increasingly dominated by compact, flat--spectrum sources 
which may be variable on timescales of a few years. 

We are currently carrying out a sensitive radio survey of the entire 
southern sky at 20\,GHz, using a wide-band analogue correlator on 
the Australia Telescope Compact Array (ATCA; see Ricci et al.\ 2004a 
for an outline of the pilot study for this survey).  We have therefore 
begun an investigation of the long--term variability of radio sources 
selected at 20\,GHz, which will also help us estimate the likely long--term 
stability of our source catalogue.  

There is little information to guide us in what to expect.  
Only a few studies of radio--source variability have been carried out 
at frequencies above 5\,GHz and these have generally targeted sources 
which were either known to be variable at lower frequencies, or were 
selected to have flat or rising radio spectra at frequencies below 
about 5\,GHz.  Such objects may not be typical of the 20\,GHz source 
population as a whole. 

The full AT 20\,GHz (AT20G) survey, using the whole 8\,GHz bandwidth 
of the analogue correlator and coherently combining all three 
interferometer baselines, began in late 2004 and has a detection 
limit of 40--50\,mJy at 20\,GHz, i.e. about a factor of two fainter 
than the sources discussed here.  It will eventually cover the entire 
southern sky from declination $0^\circ$ to $-90^\circ$. 

Our reasons for carrying out a Pilot Survey in advance 
of the full AT20G survey were to characterize 
the high--frequency radio--source population, and to optimize 
the observational techniques used in the two--step survey process 
(i.e. fast scans of large areas of sky with a wide-band analogue 
correlator, followed by snapshot imaging of candidate detections) 
to maximize the completeness, reliability and uniformity of the 
final AT20G catalogue.  Because of the slightly different 
observational techniques used in 2002, 2003 and 2004, the Pilot 
Survey data are not as complete or uniform as the AT20G data are 
intended to be.  The Pilot Survey nevertheless provides an important 
first look at the faint radio--source population at 20\,GHz. 
Since corrections for extragalactic foreground confusion will be 
critical for next--generation CMB surveys, a better knowledge of 
the properties of high--frequency radio sources (and especially 
their polarization and variability) is particularly desirable. 

This paper presents an analysis of the radio--source population 
down to a limiting flux density of about 100\,mJy at 20\,GHz, 
based on observations in the declination zone $-60^\circ$ to 
$-70^\circ$ scanned by the AT 20\,GHz Pilot Survey in 2002 and 2003.  
Our aim is to provide some first answers to the 
following questions:
\begin{itemize}
\item
How does the radio--source population at 20\,GHz relate to the 
`flat--spectrum' and `steep--spectrum' populations identified at 
lower frequencies? 
\item 
What fraction of radio sources selected at 20\,GHz are variable 
on timescales of a few years, and how stable in time is a 20\,GHz source 
catalogue? 
\item
What are the polarization properties of radio continuum sources selected 
at 20\,GHz?
\end{itemize}

\section{Observations}
\subsection{The ATCA wide--band correlator}  
An analogue correlator with 8\,GHz bandwidth (Roberts et al.\ 2006), 
originally developed for 
the Taiwanese CMB instrument AMiBA (Lo et al.\ 2001) is currently being 
used at the Australia Telescope Compact Array (ATCA) to carry out a 
radio continuum survey of the entire southern sky at 20\,GHz. 
The wide bandwidth of this correlator, combined with the fast scanning 
speed of the ATCA, makes it possible to scan large areas of sky at high 
sensitivity despite the small (2.3\,arcmin) field of view at 20\,GHz.  
Since delay tracking cannot be performed with this wide-band analogue 
correlator, all scanning observations are carried out on the meridian 
(where the delay for an east--west interferometer is zero). 
\begin{table}
\begin{center}
\begin{tabular}{lcccc}
\hline
\multicolumn{1}{c}{Date} & \multicolumn{1}{c}{N$_{\rm ant}$} & 
\multicolumn{1}{c}{$\nu_{\rm cen}$} & \multicolumn{1}{c}{Bandwidth} & \multicolumn{1}{c}{Baselines} \\
        &  & \multicolumn{1}{c}{(GHz)}  & \multicolumn{1}{c}{(GHz)}  & \multicolumn{1}{c}{(m)} \\
\hline
2002 Sep 13--17  & 2  & 18 & 3.4  &  30  \\
2003 Oct 9--16   & 3  & 17.6, 20.4 & 6--7 &  30, 30, 60 \\
\hline
\end{tabular}
\caption{Log of ATCA fast--scanning observations with the wide--band analogue 
correlator in the declination zone $-60^\circ$ to $-70^\circ$. 
N$_{\rm ant}$ is the number of antennas used for each scanning 
session. }
\label{tab:obs}
\end{center}
\end{table}

\begin{table}
\begin{center}
\begin{tabular}{lllc}
\hline
\multicolumn{1}{c}{Date} & \multicolumn{1}{c}{ATCA} & 
\multicolumn{1}{c}{Obs. Freq.} & \multicolumn{1}{c}{N$_{\rm ant}$} \\
        & \multicolumn{1}{c}{config.} & \multicolumn{1}{c}{(GHz)}  &  \\
\hline
2002 Oct 8--12  &  H168B & 17.2, 18.8             &  3  \\
2003 Nov 3--6   &  H214  & 17.0, 19.0, 21.0, 23.0 &  5  \\
2003 Nov 8--10  &  1.5D  &  4.8, 8.6              &  5  \\
2004 Oct 21-28  &  H214  & 19.0, 21.0             &  5  \\  
\hline
\end{tabular}
\caption{Log of follow--up ATCA imaging observations of sources 
detected in the scanning survey at 20\,GHz. N$_{\rm ant}$ shows the 
number of antennas equipped with 12\,mm receivers for each observing session. 
The angular resolution of the follow--up images is typically 8\,arcsec at 4.8\,GHz, 
4\,arcsec at 8.6\,GHz and 15\,arcsec at 20\,GHz. }
\label{tab:obs}
\end{center}
\end{table}

The fast-scanning survey measures approximate positions and flux densities 
for all candidate sources above the detection threshold of the survey.  
Follow-up 20\,GHz imaging of these candidate detections is then 
carried out a few weeks later, using the ATCA  in a hybrid configuration 
with its standard (delay--tracking) digital correlator.  
These follow--up images allow us to confirm detections, and to measure 
accurate positions and flux densities for the detected sources. Finally, 
the confirmed sources are also imaged at 5 and 8\,GHz to measure their 
radio spectra, polarisation and angular size. 

\subsection{Observations in the $-60^\circ$ to $-70^\circ$ declination zone} 

Tables 1 and 2 summarize the telescope and correlator configurations 
used for the observations discussed in this paper.  There are three main 
data sets: 

\begin{enumerate} 
\item
The ATCA Pilot Survey observations made in 2002 and published by 
Ricci et al.\ (2004a).  These are briefly described in \S2.4 below. 
\item
Data from a resurvey of the same declination zone at 20\,GHz in 2003, 
together with near--simultaneous observations at 4.8 and 8.6\,GHz of the  
confirmed sources (see \S2.5). 
\item
20\,GHz images made in 2004 of sources detected at 18\,GHz 
in 2002 and/or 2003, as part of a program to monitor the long--term 
variability of the sources detected in the pilot survey (\S2.6). 
\end{enumerate} 

Although our ATCA 20\,GHz pilot survey covered the whole sky between 
declinations $-60^\circ$ to $-70^\circ$, only sources with 
Galactic latitude $|b|>10^\circ$ are discussed in this paper.  
While the source population at $2<|b|<10^\circ$ is also dominated by 
extragalactic objects, it is very difficult to make optical identifications 
of radio sources close to the Galactic plane because 
of the high density of foreground stars.  Since one aim of this study 
is to examine the optical properties of high--frequency radio sources, 
we therefore chose to exclude the small number of extragalactic sources 
which lay within ten degrees of the Galactic plane, or within 5.5 
degrees of the centre of the Large Magellanic Cloud.

\subsection{The flux density scale of the ATCA at 20\,GHz}
At centimetre wavelengths, the ATCA primary flux calibrator is the 
radio galaxy PKSB\,1934--638 (Reynolds 1994).  Planets have traditionally 
been used to set the flux density scale in the 12\,mm (18--25\,GHz) band, 
and the planets Mars and Jupiter were used as primary flux calibrators 
during the first two years of operation of the ATCA 12mm receivers 
in 2002--3. However, the use of planets to set the flux density scale 
has some significant disadvantages (Sault 2003): 
\begin{itemize} 
\item
Their angular size (4--25\,arcsec for Mars and 30--48\,arcsec for Jupiter) 
means that they can be resolved out at 20\,GHz on baselines greater 
than a few hundred metres. 
\item
Their (northern) location on the ecliptic means that they are 
visible above the horizon for a much shorter time than a southern 
source like PKSB\,1934--38, and shadowing of northern sources can 
also be a problem in some compact ATCA configurations. 
\end{itemize}

PKSB\,1934--638 was monitored regularly in the 12\,mm band over a 
six--month period in 2003, using Mars as primary flux calibrator 
(Sault 2003). These observations showed that the flux density of 
PKSB\,1934--638 remained constant (varying by less than $\pm$1--2\% 
at 20\,GHz), making 
it suitable for use as a flux calibrator at these high frequencies.  
From 2004, therefore, PKSB\,1934--638 was used as the primary flux 
ATCA calibrator at 20\,GHz, whereas Mars was used in our 2002 and 
2003 observations.  

\subsection{2002 observations}
\subsubsection{Scanning observations}
The first observations of the declination strip $-60^\circ$ to $-70^\circ$ 
were made by Ricci et al.\ (2004a).  Using a single analogue 
correlator with 3\,GHz bandwidth and two ATCA antennas on a single 30\,m baseline, 
they detected 123 extragalactic ($|b|>5^\circ$) sources at 18\,GHz above a limiting 
flux density of 100\,mJy.  
The 2002 observations did not completely cover the whole $-60^\circ$ 
to $-70^\circ$ declination strip because of technical problems which 
interrupted some of the fast scanning runs. Figure 4 of Ricci et 
al.\ (2004a) shows the 2002 sky coverage and the missing regions, 
which are mainly in the RA range 5--8\,h. 
The declination $-60^\circ$ to $-70^\circ$ strip was therefore 
reobserved at 22\,GHz in 2003, and full coverage was then achieved.  
The region overlapped by the 2002 and 2003 observations gives 
a useful test of the completeness of the scanning survey technique, 
as discussed in \S4.

\subsubsection{Follow--up imaging and flux--density errors }
Follow-up synthesis imaging of the candidate sources detected 
in the 2002 scans was carried out at 18\,GHz with the ATCA as 
described by Ricci et al.\ (2004a).  It is important to note that, 
because the candidate source 
positions obtained from the wide-band scans in 2002  were typically 
accurate to $\sim$1\,arcmin, and the primary beam of the ATCA antennas 
at 20\,GHz 
is only $\sim$2.3\,arcmin, about 30\% of the sources detected in the 
follow--up images were offset by 80\,arcsec or more from the pointing 
centre, and so required large (more than a factor of two) corrections 
to their observed flux densities to correct for the attenuation of the 
primary beam.  These corrections were made by Ricci et al.\ (2004a), 
but were not explicitly discussed in their paper.  It has subsequently 
become clear that uncertainties in the primary beam correction at very 
large offsets from the field centre can sometimes introduce large systematic 
errors into the observed fluxes.  For this reason, we now regard the 
18\,GHz flux density measurements listed by Ricci et al.\ (2004a) as 
unreliable for sources observed at more than 80\,arcsec from 
the imaging field centre. For follow--up imaging in 2003 and subsequent 
years, sources more than 80\,arcsec from the imaging field centre were 
re-observed at the correct position whenever possible. 

\subsection{2003 observations}
\subsubsection{Scanning observations} 
In 2003, we used three analogue correlators and three ATCA antennas, 
giving us three independent baselines (of 30, 30 and 60\,m).  
The correlators also had a new design with the potential for 8\,GHz 
operation (Roberts et al.\ 2006). 
The 2003 fast scans were carried out using three ATCA antennas separated 
by 30\,m on an east--west baseline, and scanned in a trellis pattern at 
15\,deg\,min$^{-1}$ with 11--degree scans from declination $-59.5^\circ$ 
to $-70.5^\circ$, interleaved with 2.3\,arcmin separation and sampled at 
54\,ms.  

The system temperature was continually monitored at 17.6 and 20.4\,GHz 
and periods with high sky noise (i.e. due to clouds or rain) were flagged 
out and repeated later.  Calibration sources were observed 
approximately once per day by tracking them through transit ($\pm5$\,min).

Due to an unforeseen problem matching the wide-band receiver output to the 
fibre modulator, there was a 15\,db slope across the bandpass.   
When we transformed the 16 lag channels observed into 8 complex frequency 
channels, the resulting bandpass was uncalibratable and unphysical.  
This occurred because we had an analogue correlator and there is no exact 
Fourier Transform relation between delay and frequency (Harris \& Zmuidzinas 
2001). 

The actual bandpass was measured by taking the Fourier Transform of the 
time sequence obtained while tracking a calibrator source through transit.  
In this case we have a physical delay which changes as the earth rotates 
and we can get a sensible bandpass.  In the end only two channels 
were usable, giving a total band width of 3\,GHz.  It was also 
impractical to make a phase calibration of the three interferometers 
with this data.  As a result the sensitivity in 2003 was only marginally 
better than that in 2002, and overlapping scans could not be combined 
coherently. 

To extract a candidate source list from the 2003 raster scans, the 
correlator delays were cross-matched with the template delay pattern of 
a strong calibrator. The correlator coefficient for each time stamp along 
the scans was recorded, and values from overlapping scans were incoherently 
combined to form images in 12 equal-area zenithal projection maps 
(each two hours wide in right ascension). The source finding algorithm 
\textit{imsad} implemented in Miriad was used to extract candidate sources 
above a 5$\sigma$ threshold.

\begin{figure}
\begin{center}
\includegraphics{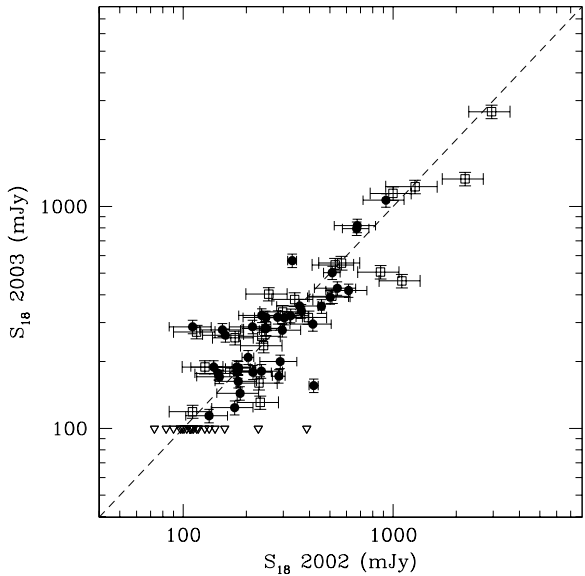} 
\vspace{8cm}
\label{fig:cumpol}
\caption{Comparison of the 18\,GHz flux densities measured in 2002 and 2003 
for sources detected independently in the scanning process.  
Sources which were detected in 2002 but not recovered in 2003 are shown as 
open triangles with a flux density limit of 100\,mJy for 2003.  
As discussed in the text, the error bars on the 2002 flux density measurements 
are significantly larger than for 2003.    
Open squares show sources with offsets of more than 80\,arcsec from the 
imaging field centre in the 2002 data. }
\end{center}
\end{figure}

\subsubsection{Follow--up imaging} 
A list of 1350 candidate sources detected in the scanning survey 
was observed at 17, 19, 21 and 23 GHz as noted in Table 2.  
As in 2002, the planet Mars was used as the primary flux calibrator.  
In the 2003 follow--up imaging, the data were reduced as the observations 
progressed, and sources which were more than 80\,arcsec from 
the imaging centre were reobserved if possible.  This significantly improved 
the accuracy of the flux density measurements for the 2003 images compared 
to those made in 2002, as can be seen in Figure 1. 

Images of each follow--up field were made at 18 and 22\,GHz using the 
multi--frequency synthesis (MFS) technique (Conway et al.\ 1990; 
Sault \& Wieringa 1994).  Since the signal-to-noise ratio in the 18\,GHz 
band was significantly higher than at 22\,GHz, we used only the 18\,GHz data 
in our subsequent analysis. 
The median rms noise in the follow--up images was 1.5\,mJy/beam at 18\,GHz, 
and  sources stronger than five times the 
rms noise level (estimated from the Stokes-V images) were considered to be 
genuine detections.  The 364 sources with confirmed detections at 18\,GHz 
(including some Galactic plane sources) were imaged at 5 and 8.6\,GHz in 
November 2003. The total integration time for these follow--up images was 
80\,s (2 cuts) at 17--19 and 21--23\,GHz, and 180\,s (6 cuts) at 5 and 
8\,GHz.   

\subsection{2004 observations}
A sample of 200 sources detected at 18\,GHz in 2002 and/or 2003 was 
re-imaged on 22 October 2004 in a series of targeted observations at 
19 and 21\,GHz, using the ATCA hybrid configuration H214.  All these 
imaging observations were centred at the source position measured in 
2002/3, so that positional offsets from the imaging field centre 
were negligible.  The 19 and 21\,GHz data were combined to produce a 
single 20\,GHz image of each target source.  The total integration 
time at 20\,GHz was 240\,s (2 cuts), and the median rms noise in the 
final images was 0.7\,mJy rms. 

\begin{figure*}
\centering
\begin{minipage}{180mm}
\vspace*{6cm}
\includegraphics{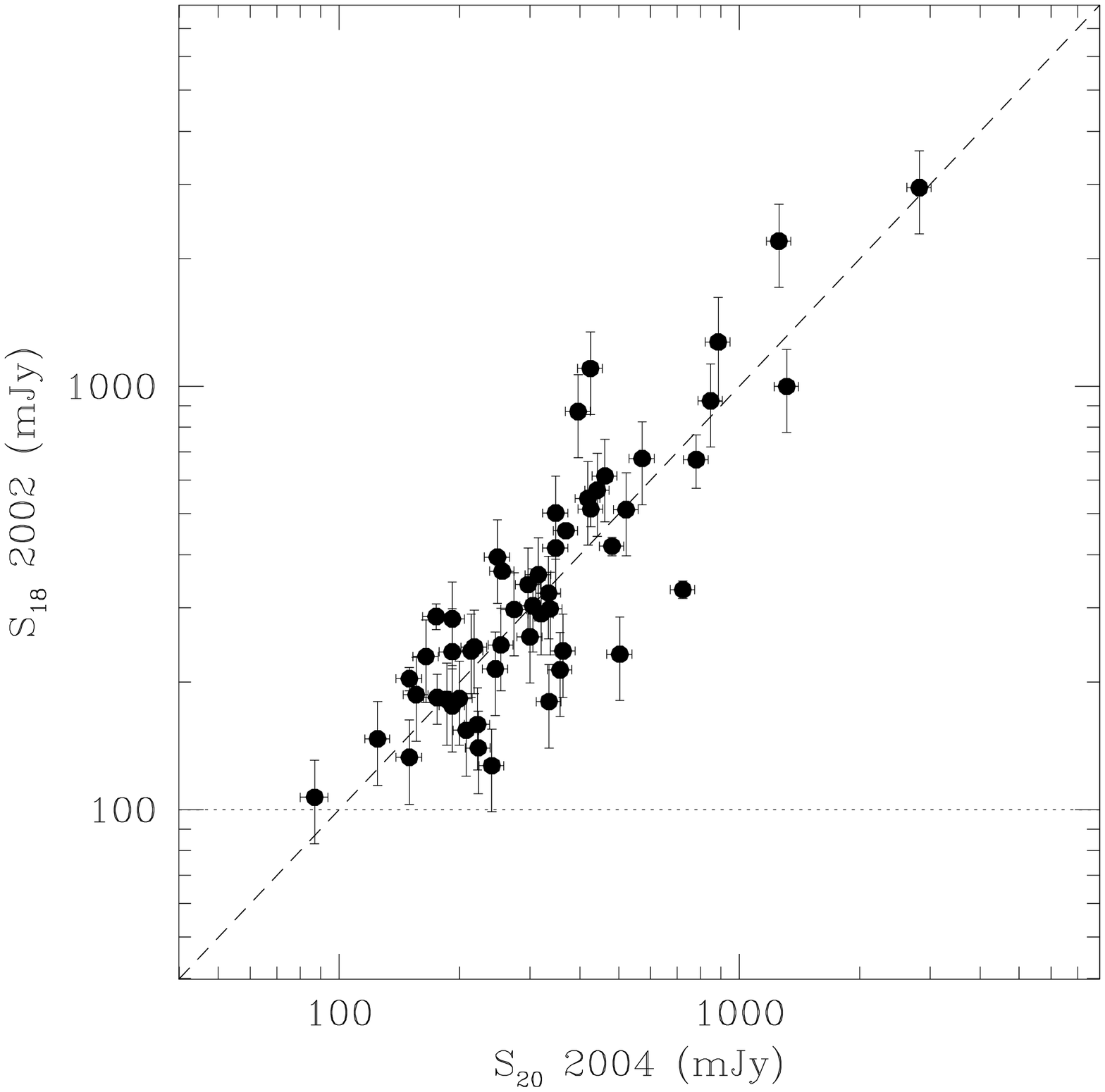}
\includegraphics{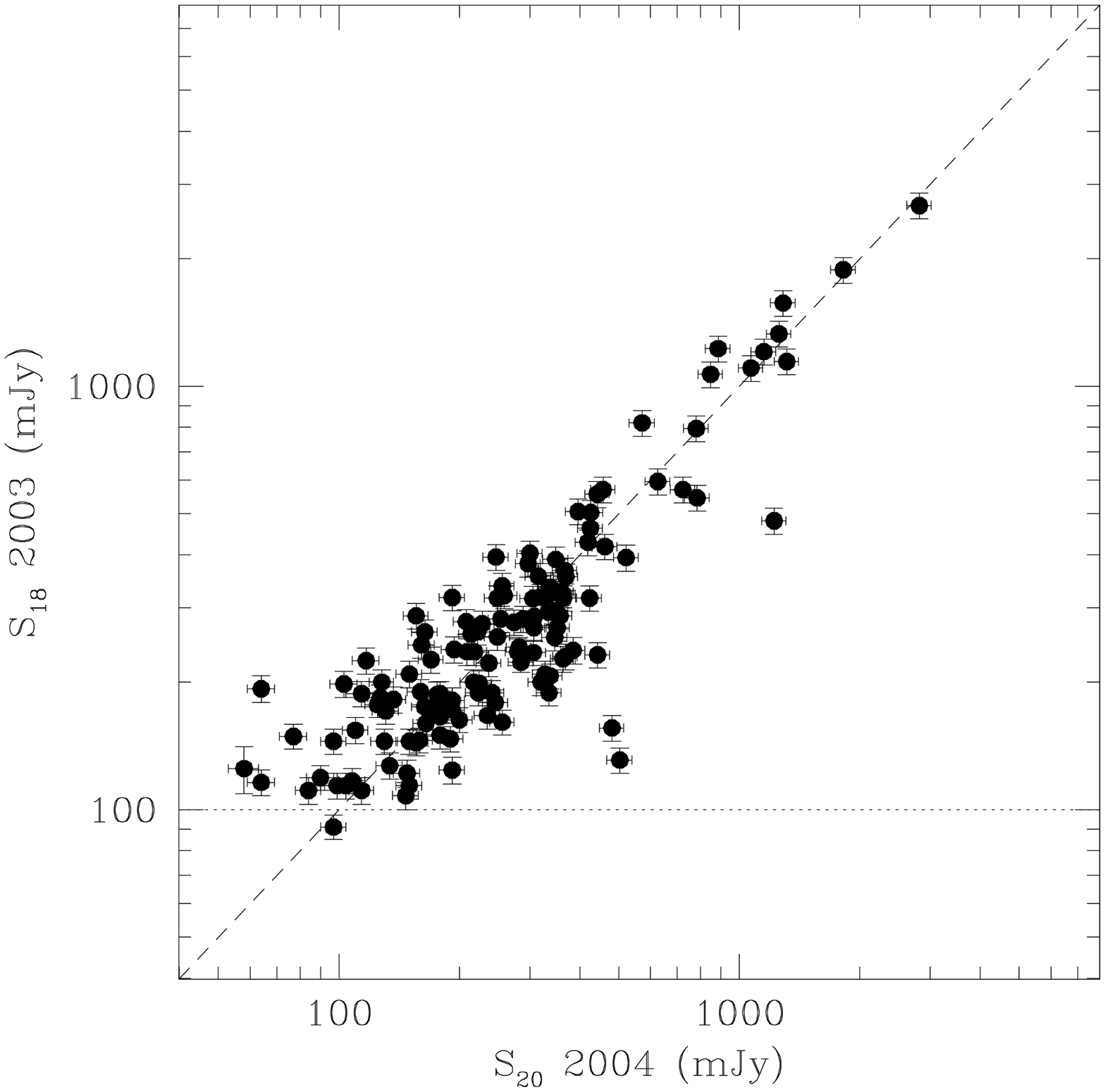}
\vspace*{1.0cm}
\caption{Comparison of 18\,GHz flux densities measured in 2002 and 2003 
with 20\,GHz flux densities measured in 2004. The horizontal dotted line 
shows the sensitivity limit of the 2002 and 2003 surveys. 
}
\end{minipage}
\end{figure*} 
 
\section{Data reduction and source--fitting }
\subsection{Reduction of the follow--up images} 
For the 2003 data, deconvolved images of the confirmed sources were made 
at 5, 8 and 18\,GHz and positions and peak flux densities were measured 
using the Miriad task \textit{maxfit}, which is optimum for a point source. 
We also used the Miriad task \textit{imfit} to measure the integrated flux 
density and angular extent of extended sources.  Where necessary, the 
fitted flux densities were then corrected for the primary 
beam attenuation at frequencies between 17 and 23 GHz based on a polynomial 
model of the Compact Array antenna pattern.

Positional errors were estimated by quadratically adding a systematic 
term and a noise term: the systematic term was assessed by cross-matching 
the 18\,GHz source positions with the Ma et al.\ (1998) International 
Coordinate Reference Frame (ICRF) source positions; the noise term is 
calculated from the synthesized beam size divided by the flux S/N. 
The median position erors are 1.3\,arcsec in right ascension and 0.6 
arcsec in declination. 

To estimate the flux density errors, we quadratically added the rms noise 
from V-Stokes images to a multiplicative gain error estimated from the 
scatter between snapshot observations of the strongest sources.  The 
median percentage gain errors were 2\% at 5 and 8\,GHz, and 5\% at 
18\,GHz.

For the 2004 data, the 19 and 21\,GHz visibilities were amplitude and phase 
calibrated in Miriad.  As noted in \S2.3, PKSB\,1934-638 was used as the 
primary flux calibrator.  The calibrated visibilities were combined to form 
20\,GHz images using the MFS technique and peak fluxes were worked out using 
the Miriad task \textit{maxfit}. Position and flux errors were determined 
in the same way as for the 2003 data. 

\subsection{Polarization measurements}
As all four Stokes parameters were available, linear polarization 
measurements were carried out on the 2003 and 2004 data. 
Q-Stokes, U-Stokes, and polarised flux $P=\sqrt(Q^2 + U^2 )$ were 
calculated at the peak of the total intensity emission at 20\,GHz.  
The rms in the V-Stokes image for each source was used as an estimate 
of the noise in U and Q.  This error estimate is then 
used to correct to first order for the Rician bias in P (Leahy 1989) 
and to set the 3$\sigma$ lower limit on P.  Note that this 
estimate corresponds to the integrated polarization for unresolved 
sources but is only the polarization at the peak in I for resolved 
sources.  Since over 95\% of the sources in our sample are unresolved 
at 20\,GHz (see \S3.4), this is not a serious problem.  

Although the measurements of fractional linear polarization made in 2003 
and 2004 were in good general agreement, the 2004 measurements had lower 
error bars and detected fractional polarizations as low as 1--2\%, whereas 
the 2003 measurements detected only the most highly--polarized sources 
with typical fractional polarizations of 4--5\% or higher.  We therefore use 
only the 2004 data in the analysis 
in \S8.2 of this paper.  A more detailed analysis of extended sources, and 
of the linear polarized flux and position angle at 5, 8 and 20\,GHz will 
be presented in a later paper.

\begin{figure}
\begin{center}
\includegraphics{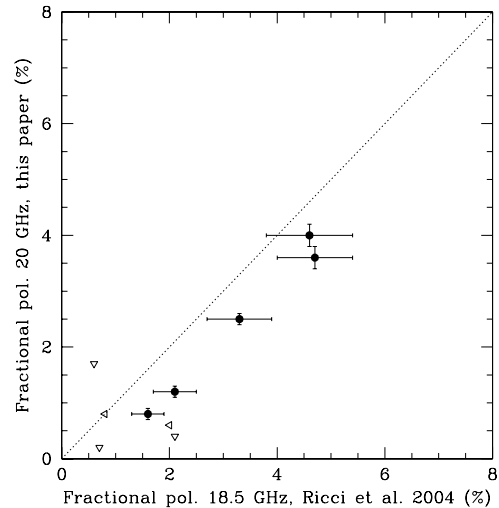} 
\vspace{7.5cm}
\caption{Comparison of the fractional linear polarization measured at 20\,GHz 
in this paper with the value measured at 18.5\,GHz by Ricci et al.\ (2004b) 
for sources in common. Filled circles show objects with polarization detected 
in both studies, and open triangles show upper limits. }
\end{center}
\end{figure}

Figure 3 compares our measurements of fractional linear polarization 
with those made by Ricci et al.\ (2004b) at 18.5\,GHz for objects 
in common.  We find a systematic difference of about 1\% in the two 
data sets, and have investigated this in consultation with authors 
of the Ricci et al.\ paper.  We find that the Ricci et al.\ (2004b) 
polarization values are about 1\% too high because these observers 
used a triple correlation method to measure polarized flux (poor phase 
stability during their run meant that they were not able to produce 
calibrated images) and did not remove the polarization bias from 
their data.  We estimate that the polarization bias in the Ricci et al.\ 
data contributes roughly 10\,mJy of spurious flux to their 
polarization measurements.  Since their sources have typical 18\,GHz 
flux densities of $\sim1$\,Jy, this corresponds to a $\sim1$\% 
increase in the measured fractional polarization compared to the 
true value. 

\subsection{The combined 20\,GHz sample 2002--2004}
Table 3 presents the observed flux densities for extragalactic sources 
with flux densities above 100\,mJy at 20\,GHz.  As noted in \S2.2, sources 
which have low Galactic latitude ($|b|<10^\circ$), or lie within 5.5 degrees 
of the centre of the Large Magellanic Cloud, have been excluded.  
A few sources with measured flux densities consistently less than 100\,mJy 
in the follow--up images were also omitted from the table.  
These sources are below the detection limit of the scanning survey and 
were found by chance in the follow--up images, 
which cover a much smaller area of sky but reach a detection limit 
of a few mJy at 20\,GHz. 

The columns in Table 3 are as follows: 
\begin{itemize}
\item[(1)] 
The AT source name, followed by \# if the source is resolved or double at 
20\,GHz (see \S3.4). 
\item[(2)]
The radio position (J2000.0) measured from the 20\,GHz images. 
For resolved doubles, the listed position is the radio centroid. 
\item[(3)] 
For sources where we were able to make an optical identification on the 
Digitized Sky Survey, this column lists the b$_{\rm J}$ magnitude from the 
Supercosmos database. 
\item[(4)] 
The object type of the optical ID, as classified in Supercosmos: T=1 for a galaxy, 
T=2 for a stellar object (QSO candidate). T=0 indicates either a blank field 
at the source position or a faint ($>$22\,mag) object for which the Supercosmos 
star/galaxy separation is unreliable. 
\item[(5)]  
The 18\,GHz flux density measured in 2002, followed by its error.  For resolved 
doubles, we list the integrated flux density over the source. Flux densities in 
square brackets\ [\ ] are measurements made at offsets of more than 80\,arcsec 
from the imaging field centre at 18--20\,GHz, and should be regarded as unreliable 
because of the large primary--beam correction . Flux densities followed by a colon 
are measured at offsets of 60--80\,arcsec from the field centre, but should be reliable. 
\item[(7)] 
The 18\,GHz flux density measured in 2003, and its error.  
\item[(9)]  
The 20\,GHz flux density measured in 2004, and its error.  
\item[(11)] 
The 8.6\,GHz flux density measured in 2003, and its error.  
\item[(13)]  
The 4.8\,GHz flux density measured in 2003, and its error.   
\item[(15)] 
The integrated flux density at 843\,MHz and its error, from the Sydney University 
Molonglo Sky Survey (SUMSS) catalogue (Mauch et al.\ 2003).  
\item[(17)]  
The fractional linear polarization at 20\,GHz measured in 2004, and its error.  
\item[(19)]
The debiased variability index at 20\,GHz, calculated as described in \S5.1. 
\item[(20)] 
Alternative source name, from the NASA Extragalactic Database. 
\item[(21)] 
Notes on individual sources, coded as follows: \\
C = listed in the online ATCA calibrator catalogue, \\
E = possible EGRET gamma--ray source (Tornikoski et al. 2002), \\
I = listed as an IRAS galaxy in the online NASA Extragalactic Database (NED), \\
M = galaxy detected in the near-infrared Two-Micron All-Sky Survey (2MASS), \\
P = in the Parkes quarter-Jy sample (Jackson et al. 2002), \\
Q = listed as a QSO in NED, \\
V = VLBI observation with the VSOP satellite (Hirabayashi et al.\ 2000) \\
W = source detected in the first--year WMAP data (Bennett et al. 2003),\\
X = listed as an X-ray source in NED, \\
* = polarization observation by Ricci et al.\ (2004b). 
\end{itemize}

\subsection{Extended sources at 20\,GHz}
The great majority of the sources detected in the 20\,GHz Pilot Survey 
are unresolved in our follow--up images at 5, 8 and 20\,GHz. The source--detection 
algorithm used in the Pilot Survey was optimized for point sources, and there will be 
some bias against extended sources with angular sizes larger than about 30\,arcsec.  
For sources larger than 1\,arcmin in size, the total flux densities listed in Table 3 
may also be underestimated. 

Only eleven of the 173 sources in Table 3 were resolved 
in our (15\,arcsec resolution) 20\,GHz images. 
The overall properties of extended sources in the current sample are as 
follows:
\begin{itemize}
\item
Three objects, J0103--6439, J2157--6941 and J2358--6052/J2359--6057, 
are very extended double--lobed radio sources which are too large to be imaged 
with these ATCA snapshots.  As a result, the total flux densities listed in 
Table 3 are lower limits to the correct value. 
\item 
Another seven sources are resolved in our ATCA images, but still lie 
within the 2.2\,arcmin primary beam of the ATCA at 20\,GHz.  
Details of these objects are given in Appendix A. 
\item
Five of the extended sources (J0103--6439, J0121-6309, J0257-6112, J0743--6726 and 
J2157--6941) have a flat--spectrum core which dominates the flux density at 
20\,GHz. 
\end{itemize}
Since the number of extended sources is small, and they appear to be 
somewhat diverse in nature, we defer any detailed discussion of the 
extended radio--source population to a later paper.  

\subsection{Optical identification of the 20\,GHz sources} 

\begin{figure}
\begin{center}
\includegraphics{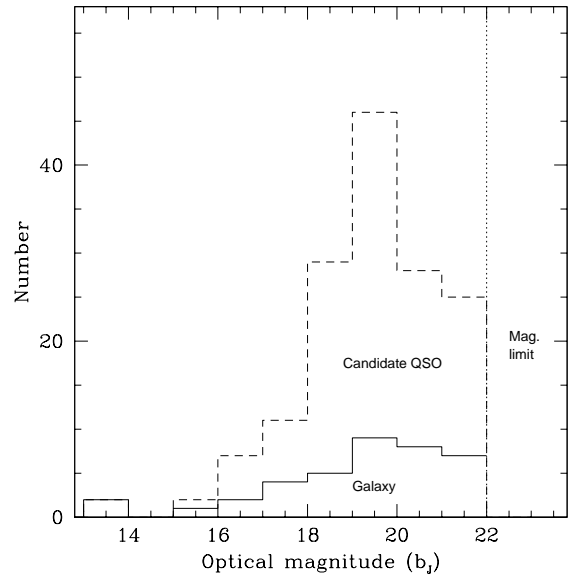} 
\vspace{8cm}
\label{fig:opt_id}
\caption{Optical identifications for the 20\,GHz radio sources in Table 3.  
Galaxies and stellar objects (QSO candidates) are shown separately. 
Only 27 sources (13\% of the sample) are unidentified down to 
b$_{\rm J}<$22\,mag. }
\end{center}
\end{figure}

\begin{figure}
\begin{center}
\includegraphics{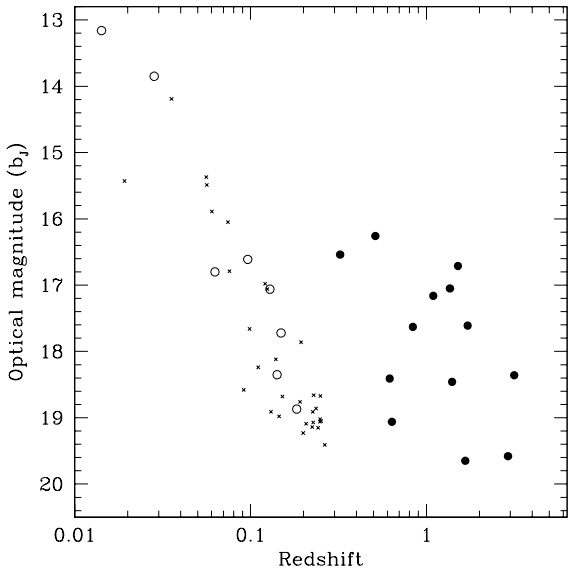} 
\vspace{8cm}
\label{fig:z_mag}
\caption{Relation between SuperCOSMOS $b_{\rm J}$ magnitude and redshift for those 
objects in our sample which have a published redshift. Open circles show 
galaxies from the 20\,GHz and filled circles QSOs.  The small crosses 
show a representative subsample of 2dFGRS radio galaxies selected at 
1.4\,GHz (Sadler et al.\ 2002).  The highest redshift so far measured for 
an object in this sample is for J1940--6907, a QSO at $z$=3.154.  }
\end{center}
\end{figure}

We examined all the sources in Table 3 in the SuperCOSMOS online 
catalogue and images (Hambly et al.\ 2001).  An optical object was 
accepted as the correct ID for a 20\,GHz radio source if it was brighter 
than $b_{\rm J}$ = 22\,mag.\ and lay within 2.5\,arcsec 
of the radio position.  For one source in Table 3 (J0715--6829) the optical 
image was saturated by light from a nearby 11th magnitude star and so no 
identification could be attempted. Of the remaining 172 sources, 
146 (85\%) had an optical ID which met the criteria listed above.  
Monte Carlo tests (based on matching the SuperCOSMOS catalogue with 
radio positions randomly offset from those in Table 3) imply that at 
least 97\% of these IDs are likely to be genuine associations, rather 
than a chance alignment with a foreground or background object. 

As can be seen from Figure 4, the majority (65\%) of radio sources selected 
at 20\,GHz have stellar IDs on the DSS B images, and are candidate QSOs 
or BL Lac objects.  20\% of the radio sample are identified with galaxies and 
15\% are faint objects or blank fields.  The overall optical identification 
rate of 85\% for radio sources selected at 20\,GHz is significantly 
higher than the identification rate for bright radio sources selected at 
1.4\,GHz (typically $\sim$30\% above B$\sim$22\,mag), but is closer  
to that found by Bolton et al.\ (2004) for a flux--limited sample of 
radio sources selected at 15\,GHz, as discussed in \S8.1.1. 

Figure 5 shows the relation between $b_{\rm J}$ magnitude and redshift 
for the 22 sources (13\% of the objects in Figure 4) which currently 
have a published redshift.  A representative sample of nearby radio 
galaxies (Sadler et al.\ 2002) selected from the 2dF Galaxy Redshift 
Survey (2dFGRS; Colless et al.\ 2001) is shown for comparison.  
Galaxies detected in our 20\,GHz survey appear to 
span a narrow range in optical luminosity similar to 
that seen in nearby radio galaxies selected at lower frequencies, 
though we caution that the sub-sample of sources with published 
redshifts is inhomogeneous in nature and may be biased in luminosity 
and/or redshift distribution because redshifts are easier to measure for 
brighter galaxies and QSOs. 

\section{Reliability and reproducibility of the scanning survey } 
As noted earlier, the fact that our pilot survey scanned the same 
area of sky in both 2002 and 2003 provides an important test of the  
observational techniques to be used for the full AT20G survey 
(i.e. fast scans of large areas of sky, followed by imaging of 
candidate sources identified in the scans).  In particular, 
how good a job does the scanning technique do in finding genuine 
sources down to the nominal detection limit, and how reproducible 
are the source lists produced by this technique? 

Table 4 shows the recovery rate in 2003 of sources detected in the 
2002 Pilot Survey scans.  As might be expected (since the nominal 
detection limit of the 2003 survey was 100\,mJy), none of the weakest 
(S$_{18}<100$\,mJy) sources detected in 2002 was recovered in the 2003 
scans, but the recovery rate rises to 95\% for sources 
with measured flux densities above 150\,mJy in 2002.   
We checked the three sources above 150\,mJy which were not recovered in 2003.  
In all cases these sources were visible in the raw data scans, so had not decreased 
in flux density to below the survey limit.  Instead, they were missed from the 
follow--up imaging program because of deficiencies in the source--detection algorithm 
for extended sources or sources with nearby bad data points. For the full AT20G survey, 
we will use an improved source--detection algorithm. 

\begin{table}
\setcounter{table}{3}
\begin{center}
\begin{tabular}{lrrr}
\hline
\multicolumn{1}{c}{S$_{18}$} & \multicolumn{1}{c}{Observed }  & \multicolumn{1}{c}{Recovered } & 
\multicolumn{1}{c}{Fraction } \\
\multicolumn{1}{c}{(mJy)  } & \multicolumn{1}{c}{in 2002} & \multicolumn{1}{c}{in 2003} 
& \multicolumn{1}{c}{recovered} \\
 \hline
  $<$100       &   6  &   0   &   0\%  \\
  101--125     &  13  &   4   &  31\%  \\ 
  126--150     &   8  &   5   &  63\%  \\
  151--200     &  12  &  12   & 100\%  \\
  $>$200       &  50  &  47   &  94\%  \\ 
\hline
\end{tabular}
\caption{The fraction of sources detected at 18\,GHz in the 2002 scans which 
were independently detected in the 2003 scans of the Pilot Survey area.  }
\end{center}
\end{table}

Figure 1 compares the 18\,GHz flux densities measured in 2002 and 2003 
for sources detected in both years.  It implies that the flux density 
scales are in good agreement, and gives some first hints that the 
general variability level at 18\,GHz is modest (though the large 
error bars on the 2002 flux densities mean that this is not a very  
useful data set for studying variability in a quantitative way). 
We therefore conclude that the scanning technique produces a 
reliable and robust catalogue of sources, in the sense that re-scanning 
an area of sky will produce essentially the same source catalogue each 
time.

\begin{figure*}
\centering
\begin{minipage}{180mm}
\vspace*{4.6cm}
\includegraphics{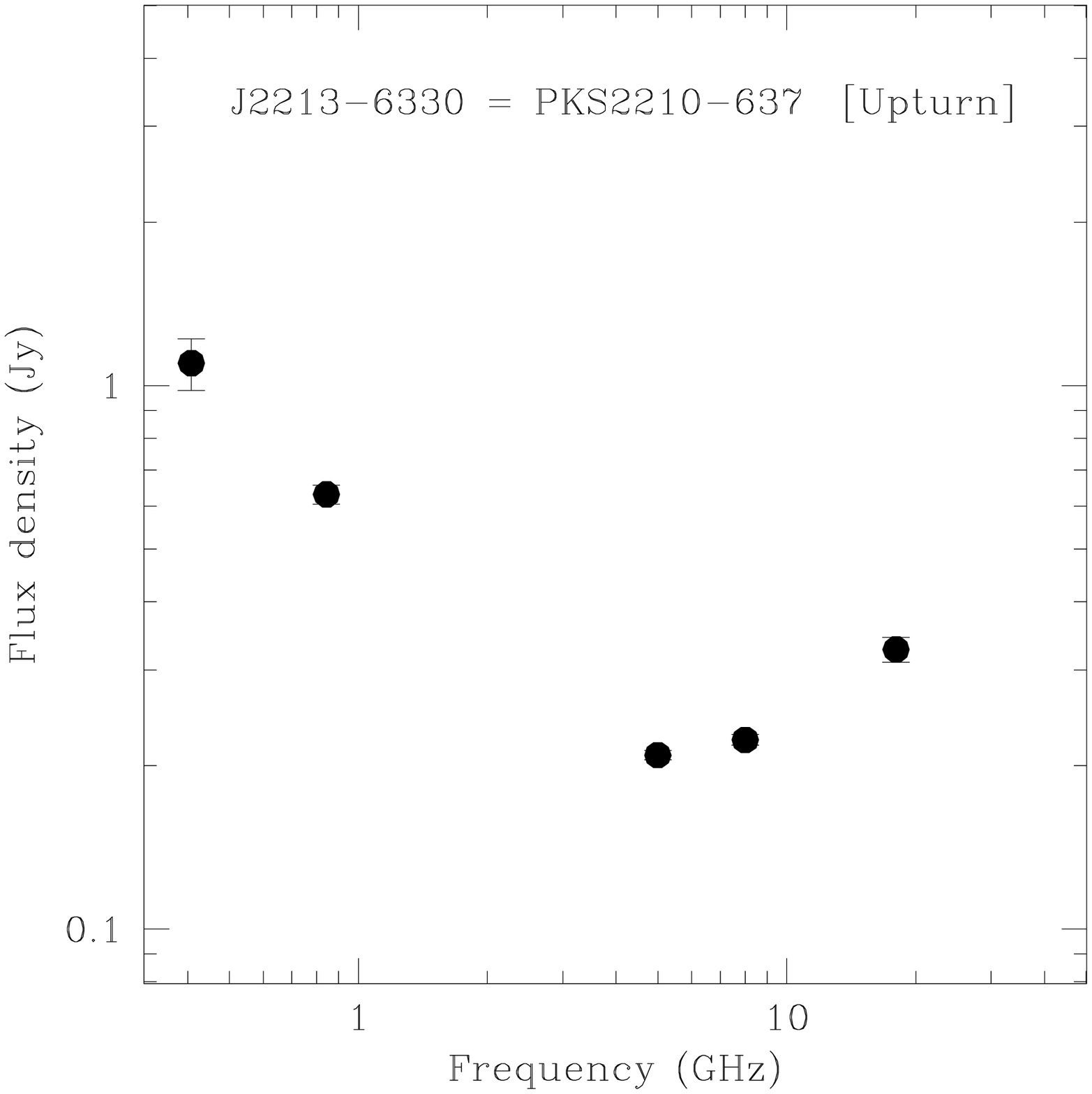}
\includegraphics{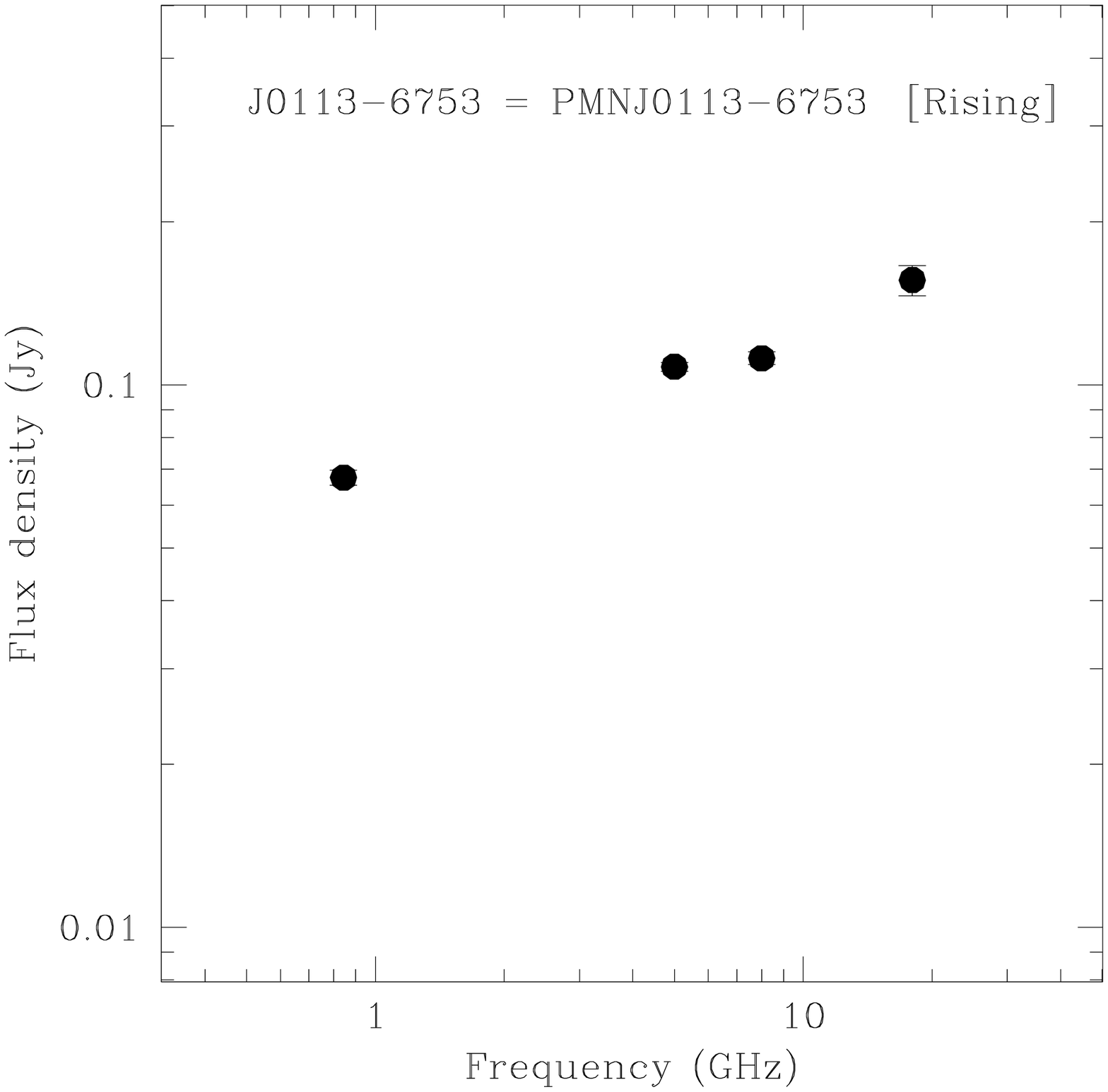}
\includegraphics{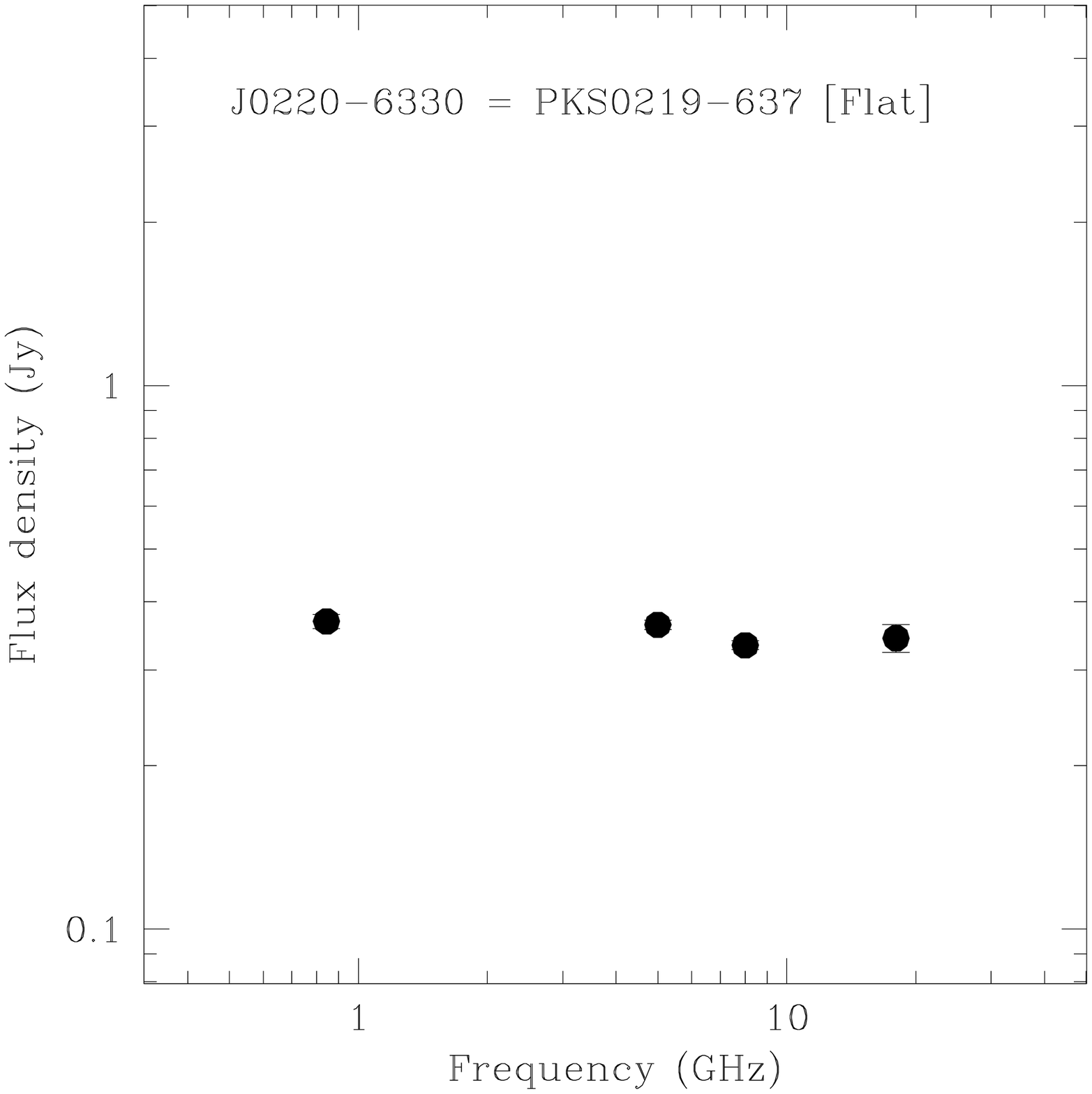}
\vspace*{6cm}
\includegraphics{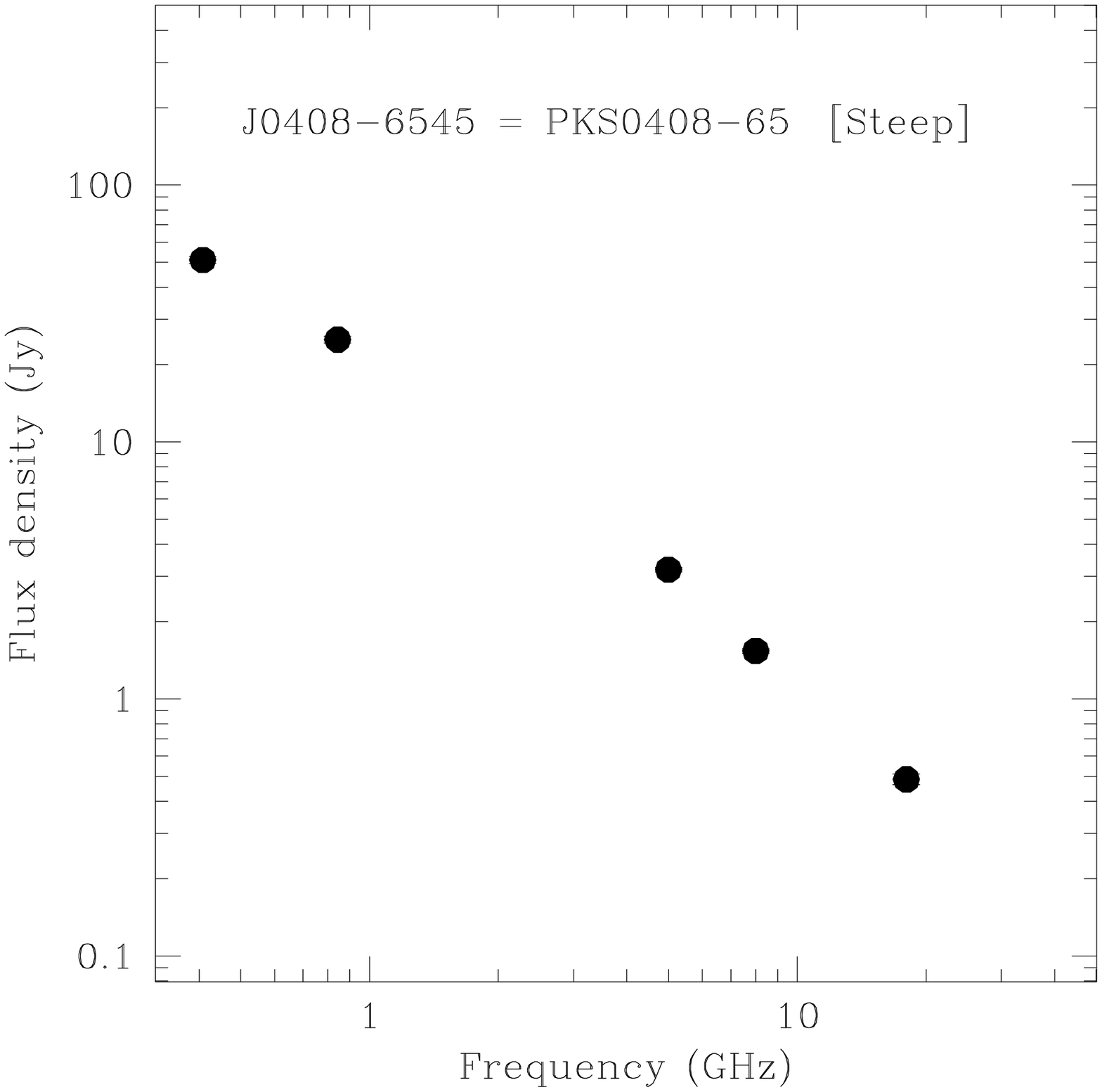}
\includegraphics{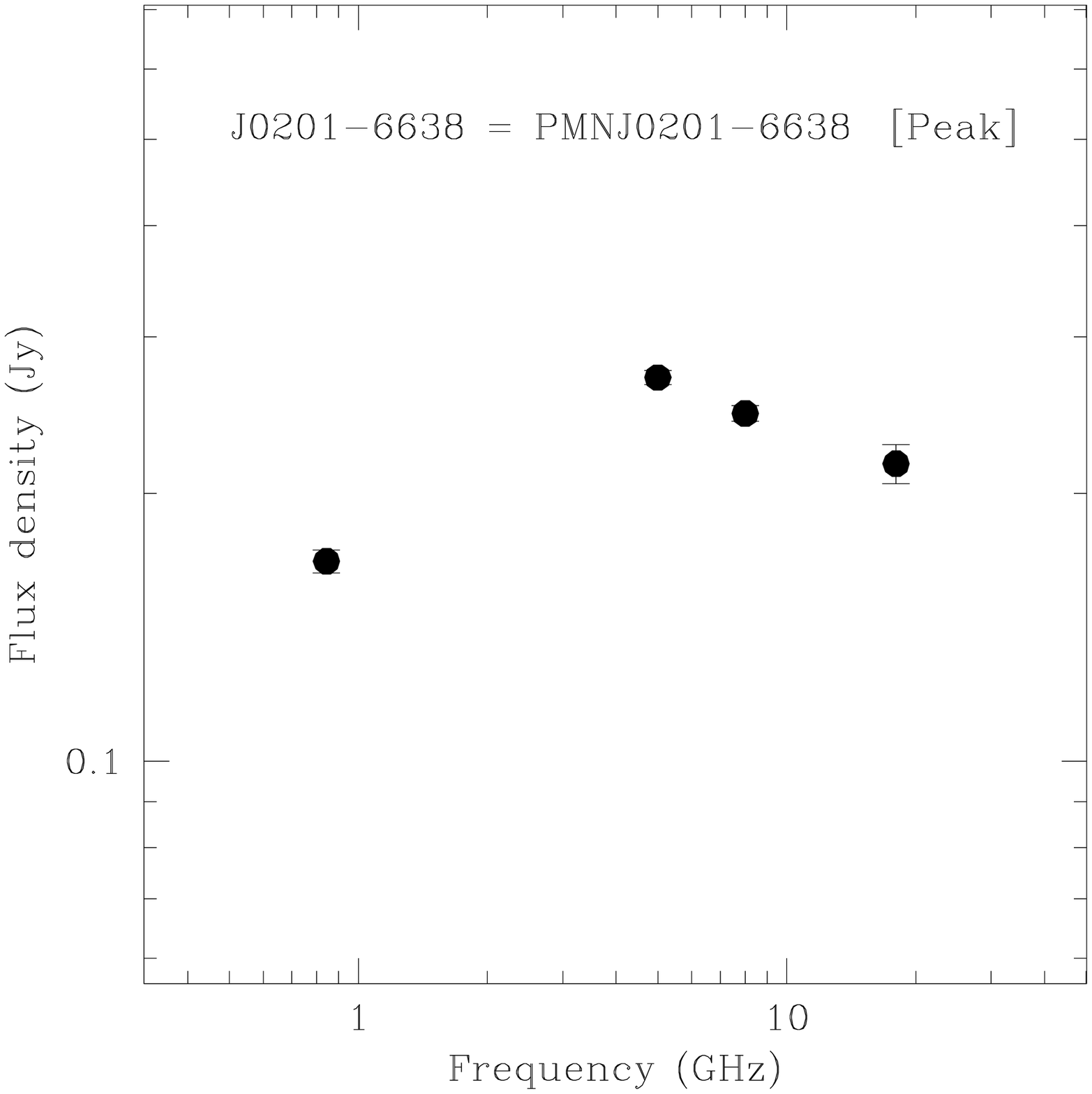}
\vspace*{1.0cm}
\caption{Examples of radio spectra for each of the four spectral classes 
identified in the text (Upturn, Rising, Steep and Peak), together with a 
spectrum classified as Flat ($|\alpha|<0.1$ for both 0.84--5\,GHz and 
8--20\,GHz). Where available, a 408\,MHz flux density from the MRC 
(Large et al.\ 1981) is plotted in addition to the data from Table 3.  } 
\end{minipage}
\end{figure*} 

\section{Radio spectra of the 20\,GHz sources}
\subsection{Representative radio spectra at 0.8 to 20\,GHz} 
Figure 6 shows some representative radio spectra for sources in our 
sample.  It is clear we see a wide variety of spectral shapes, most 
of which cannot 
be fitted by a single power--law over the frequency range 1--20\,GHz.  
We can distinguish four main kinds of spectra: \\

\noindent
(a) Sources with steep (falling) spectra over the
    whole range 843 MHz to 20 GHz (e.g. J0408--6545 in Fig.\ 6). \\
(b) Sources with peaked (GPS) spectra, in which the flux density
    rises at low frequency and falls at high frequency (e.g. J0201--6638). \\
(c) Sources with inverted (rising) radio spectra over the whole frequency
    range (e.g. J0113--6753). \\
(d) Sources with an upturn in their spectrum, where the flux density is 
falling at lower frequencies, but then turns up and begins to rise above 5--8\,GHz 
(e.g. J2213--6330).  \\ 

In addition, a small number of sources have flat radio spectra in which 
the flux density is essentially constant over the entire frequency range 
observed (e.g. J0220--6330 in Fig.\ 6). \\

The radio spectral index 
$$\alpha=\frac{({\rm log} S_1-{\rm log} S_2)}{({\rm log} \nu_1-{\rm log} \nu_2)}$$ 
where $S_1$ and $S_2$ are the measured flux densities at frequencies $\nu_1$ and $\nu_2$,  
is commonly used to characterize radio--source populations at centimetre wavelengths 
(frequencies of 1 to 5 GHz) where many large--area radio surveys have been carried out. 

At centimetre wavelengths the radio emission from flat--spectrum ($\alpha$\,$>$\,$-0.5$) 
objects is dominated by a compact, self-absorbed component, while steep--spectrum 
objects ($\alpha$\,$<$\,$-0.5$) are dominated by optically--thin synchrotron emission.  
The flat-- and steep--spectrum populations are usually considered separately when 
modelling the cosmic evolution of radio sources (e.g. Peacock \& Gull 1981; 
Peacock 1985). 

As pointed out by Peacock (1985), the radio spectral index is only valid as a 
diagnostic tool if it is measured over a frequency interval small enough that the effects 
of spectral curvature can be neglected.  Because many of the sources in our sample 
have significant spectra curvature over the frequency range 1--20\,GHz, we therefore 
use a `radio two--colour' diagram, rather than a single spectral index, to characterize 
the high--frequency radio--source population.  As discussed in the next section, 
this diagram compares a low--frequency 
spectral index $\alpha_L$ (which corresponds closely to the spectral index 
traditionally used to separate flat-- and steep--spectrum radio sources) with a 
high--frequency spectral index $\alpha_H$ which measures the spectral shape 
above 8\,GHz. 

\subsection{The radio two--colour diagram}
Figure 7 shows the radio two--colour diagram, which compares the spectral 
indices $\alpha_L$ (at 0.84--5\,GHz) and $\alpha_H$ (at 8--18\,GHz) for sources 
selected at 20\,GHz. 
Such a diagram is analogous to the two--colour 
diagram used in optical astronomy to characterise the broad--band continuum properties 
of stars and galaxies.  The diagram shown in Figure 7 has the advantage that 
the axes (and error bars) are independent, and the four main spectral classes identified 
in Figure 6 correspond to the four quadrants in the two-colour diagram. 
The dotted line shows the relation for galaxies whose spectra follow 
a single power--law from 0.8 to 18\,GHz.  Only a small 
fraction of sources fall on or near the dotted line, and it is clear 
that $\alpha_L$ and $\alpha_H$ are not strongly correlated. 

\begin{figure}
\begin{center}
\special{
psfile=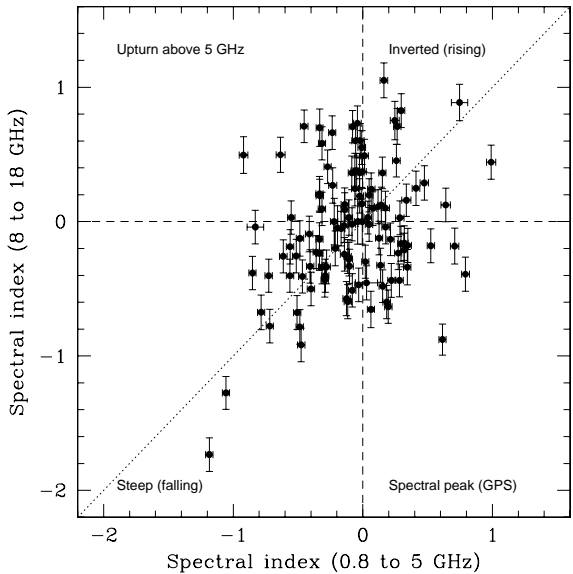
hscale=40 vscale=40 hoffset=0 voffset=-280} 
\vspace{8cm}
\label{fig:cumpol}
\caption{Radio `two--colour diagram' for the 119 extragalactic sources 
in Table 3 which have good--quality multi--frequency observations made 
in late 2003.  }
\end{center}
\end{figure}

\begin{figure}
\begin{center}
\includegraphics{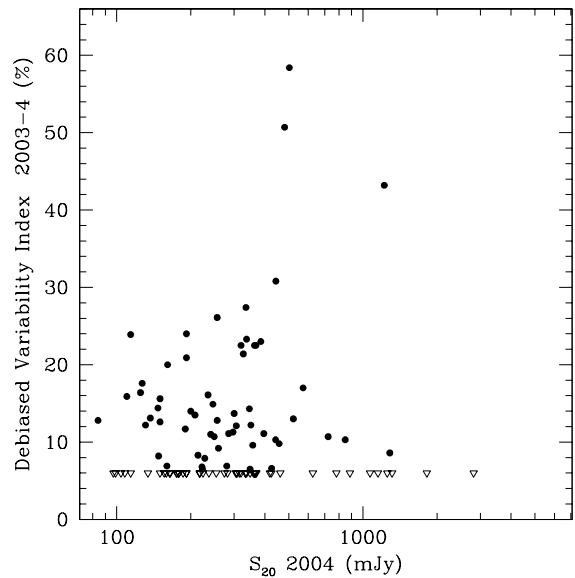} 
\vspace{7.5cm}
\caption{Debiased variability index at 20\,GHz, measured from 2003--4 for 
121 sources observed at both epochs. Variability of 6\% or more over this 
one--year interval is detectable in the current data set. }
\end{center}
\end{figure}

Over 30\% of the sources in Figure 7 have flat or
inverted spectra between 8 and 20 GHz (i.e. $\alpha_H$\,$>$\,0), 
and more than half of these 
(like J2213-6330  in Fig.\ 6) have steep radio 
spectra below 5\,GHz and would not have been predicted as 
strong 20\,GHz sources 
on the basis of their low--frequency spectra.  In contrast, many sources with 
flat or inverted spectra below 5\,GHz turn over and become steep above 8\,GHz.  

\begin{table}
\begin{center}
\begin{tabular}{lccc}
\hline
\multicolumn{1}{l}{Spectrum} & \multicolumn{1}{c}{How defined} &\multicolumn{1}{c}{Number} & 
\multicolumn{1}{c}{Fraction} \\
 \hline
Steep        &  $\alpha_L$\,$<$\,0, $\alpha_H$\,$<$\,0   &  32  &  $32\pm6$\%   \\
Upturn       &  $\alpha_L$\,$<$\,0, $\alpha_H$\,$>$\,0   &  22  &  $22\pm5$\%   \\
Rising (Inv.)&  $\alpha_L$\,$>$\,0, $\alpha_H$\,$>$\,0   &  18  &  $18\pm4$\%   \\
Peak         &  $\alpha_L$\,$>$\,0, $\alpha_H$\,$<$\,0   &  23  &  $23\pm4$\%   \\
Flat         &  $-0.1$\,$<$\,($\alpha_L, \alpha_H$)\,$<$\,0.1 & 6  &  $6\pm2$\%   \\
\hline
Total        &                                            & 101  &      \\
\hline
\end{tabular}
\caption{Distribution of our sample in the five spectral classes defined in \S4.1 and Figure 6. }
\end{center}
\end{table}

\begin{table}
\begin{center}
\begin{tabular}{rrl}
\hline
\multicolumn{1}{c}{Debiased} & \multicolumn{1}{c}{n} & 
\multicolumn{1}{c}{Fraction}  \\
\multicolumn{1}{c}{Variability} &  &  \\
\hline
$<10$\%  &  63 &  $58\pm7$\%  \\
10--20\% &  29 &  $27\pm5$\%   \\
20--30\% &  11 &  $10\pm3$\%   \\
$>$30\%  &   5 &  $ 5\pm2$\%   \\
 Total   & 108 &    \\
\hline
\end{tabular}
\caption{Distribution of the debiased variability index at 20\,GHz for 
radio sources in our sample. }
\label{tab:obs}
\end{center}
\end{table}

\section{Variability at 20\,GHz} 
\subsection{Previous work}
As mentioned in the introduction, there have been only a few studies of 
radio--source variability at frequencies above 5\,GHz. 
Owen, Spangler \& Cotton (1980) investigated the variability of a sample 
of strong (S$_{90}>1$\,Jy) flat--spectrum sources at 5 and 90\,GHz over a 
one--year period.  They found that these sources were only slightly more 
variable at 90\,GHz than at 5\,GHz, in contrast to what they had expected.
%
Tingay et al.\ (2003) also found that the level of variability of strong, 
compact radio sources increased only moderately with frequency.  They used 
the ATCA to monitor a sample of 202 sources 
from the VSOP all--sky survey at 1.4, 2.5, 4.8 and 8.6\,GHz, and 
found a median variability of 6\% at 1.4\,GHz and 9\% at 8.6\,GHz over a 
timescale of 3--4 years.  Barvainis et al.\ (2005) found similar variability 
levels at 8.5\,GHz for a sample of radio--loud and radio--quiet QSOs observed 
at the VLA. 

These studies are based largely on objects pre-selected at lower 
frequencies, and may not give a complete picture of the variability of the 
20\,GHz source population as a whole. 

\subsection{Quantifying variability} 
Following Barvainis et at al.\ (2005) and Akritas \& Bershady (1996), 
we use a debiased variability index which takes into account the 
uncertainties in individual flux density measurements. We define 
the fractional variability index, V$_{\rm rms}$ by 
$$  {\rm V}_{\rm rms}=\frac{100}{\langle S\rangle}\sqrt{\frac{\sum[S_i-\langle S\rangle]^2-\sum\sigma_i^2}{N} }$$
where S$_i$ are individual flux density measurements for the same source, 
$\sigma_i$ is the error on each measurement, N is the number of data points, 
and $\langle S\rangle$ is the mean flux density. We follow Barvainis et 
al.\ (2005) in setting the variability index to be negative when the value 
inside the square root becomes negative.  However rather than listing negative 
values of the variability index, as was done by Barvainis et al., 
we used the distribution of positive and negative 
values to define the minimum level of variability which is detectable in 
our data.  This was found to be 6\%, so for all sources with a variability 
index below this we list the debiased variability index as $<$6.0\%. 

We used only the 2003 and 2004 data sets in our variability analysis, 
since these have significantly smaller flux--density errors than the 
2002 data. 
The 2003 imaging observations were made at 18\,GHz and the 2004 
observations at 20\,GHz, so there is a possibility of measuring 
spurious `variability' for sources which have steeply--falling radio 
spectra at 18--20\,GHz. To overcome this, we used the 8--18\,GHz spectral 
index from Table 3 to extrapolate the 2003 measurements to 20\,GHz for 
the small number of sources in Table 3 which have $\alpha<-0.7$ between 
8 and 18\,GHz.  
As discussed in \S2.3, different ATCA primary flux calibrators were 
also used for the 2003 and 2004 observing runs.  Since there appears 
to be no systematic offset between our 2003 and 2004 flux 
density measurements at 20\,GHz (see Figure 2), we do not believe 
that this has introduced any spurious 'variability' into our data set. 

Figure 8 plots the debiased variability index against flux density. 
Although several of the strongest sources in our sample are also highly 
variable (in particular 0623$-$6436, 1546$-$6837 and 1903$-$6749 in Table 3), 
there appears to be no strong correlation between variability and flux 
density for the sample as a whole. For example, the generalized Kendall's 
tau correlation implemented in the ASURV statistical package for censored data 
(Isobe, Feigelson \& Nelson 1986) has a value of 1.66, corresponding to a 
probability of 9.7\% that no correlation is present.  
Table 6 shows the distribution of the debiased variability index measured 
for sources in Table 3 over a one--year timespan from October 2003 
to October 2004.  The majority of sources (58\%) vary by less than 
10\% in 20\,GHz flux density over this period, and only five sources 
varied by more than 30\%. The median debiased variability index at 20\,GHz 
is 6.9\%.

\begin{table*}
\centering
\begin{minipage}{170mm} 
\begin{tabular}{@{}lclllllllrr@{}}
\hline
\multicolumn{1}{c}{Name} & WMAP & \multicolumn{1}{c}{S$_{20}$} &  \multicolumn{1}{c}{S$_{23}$} & \multicolumn{1}{c}{S$_{33}$} & 
\multicolumn{1}{c}{S$_{41}$} & \multicolumn{1}{c}{S$_{61}$} & \multicolumn{1}{c}{$\alpha_{\rm H}$} & 
\multicolumn{1}{c}{$\alpha_{\rm WMAP}$} &  \multicolumn{1}{c}{Pol.} &  \multicolumn{1}{c}{Var}\\
  & \multicolumn{1}{c}{cat.\ no.} & \multicolumn{1}{c}{ATCA (Jy)} & \multicolumn{4}{c}{ ----- WMAP (Jy) ----- } & & & \multicolumn{1}{c}{\%} & \multicolumn{1}{c}{\%} \\
\hline
0303--6211 & 162 & $1.26\pm0.09$ & $1.5\pm0.1$ & $1.6\pm0.2$ & $1.5\pm0.2$ & $1.7\pm0.3$ & $-0.51\pm0.12$ & $+0.1\pm0.4$ &   2.5 & $<$6.0 \\
0309--6058 & 160 & $1.15\pm0.08$ & $1.3\pm0.1$ & \nodata     & \nodata     & $1.6\pm0.5$ & $-0.04\pm0.12$ & $+0.2\pm0.8$ &   0.8 & $<$6.0 \\
0506--6109 & 154 & $1.82\pm0.12$ & $2.9\pm0.09$& $2.4\pm0.1$ & $2.0\pm0.2$ & $1.5\pm0.3$ & $+0.09\pm0.12$ & $-0.4\pm0.3$ &   1.2 & $<$6.0 \\
0546--6415 & 156 & $0.36\pm0.03$ & $0.8\pm0.06$& $0.6\pm0.09$& $0.6\pm0.1$ & \nodata     & $+0.25\pm0.13$ & $-0.6\pm0.8$ &$<$0.6 &   22.5 \\
0743--6726 & 161 & $1.07\pm0.08$ & $1.6\pm0.1$ & $1.0\pm0.1$ & $1.0\pm0.2$ & \nodata     & $-0.38\pm0.12$ & $-1.1\pm0.6$ &   4.8 & $<$6.0 \\
1703--6212 & 198 & $1.29\pm0.09$ & $1.9\pm0.1$ & $2.0\pm0.2$ & $2.3\pm0.2$ & $2.0\pm0.3$ & $+0.83\pm0.13$ & $+0.1\pm0.3$ &   0.5 & $<$6.0 \\
1723--6500 & 196 & $2.82\pm0.20$ & $2.1\pm0.1$ & $2.0\pm0.2$ & $1.6\pm0.2$ & \nodata     & $+0.02\pm0.12$ & $-0.3\pm0.4$ &$<$1.7 & $<$6.0 \\
1803--6507 & 199 & $1.31\pm0.09$ & $1.2\pm0.1$ & $1.4\pm0.2$ & $1.5\pm0.2$ & $1.2\pm0.3$ & $+0.53\pm0.12$ & $+0.2\pm0.5$ &   0.8 & $<$6.0 \\
1819--6345 & 200 & $1.83\pm0.13$ & $1.7\pm0.1$ & $1.2\pm0.2$ & $1.1\pm0.2$ & $1.4\pm0.5$ & \nodata        & $-0.7\pm0.5$ &   0.8 &\nodata \\
2035--6846 & 194 & $0.46\pm0.03$ & $1.0\pm0.2$ & $1.3\pm0.2$ & $0.9\pm0.2$ & \nodata     & $-0.18\pm0.13$ & $+0.4\pm0.6$ &   4.2 &   9.8 \\
2157--6941 & 190 & $1.99\pm0.11$ & $3.6\pm0.1$ & $2.9\pm0.2$ & $2.6\pm0.2$ & $2.2\pm0.4$ & \nodata        & $-0.6\pm0.2$ &   2.4 &\nodata \\
2359--6052 & 187 & $>$0.9        & $1.8\pm0.1$ & \nodata     & $1.3\pm0.1$ & \nodata     & \nodata        & $-0.5\pm0.5$ &   5.2 &\nodata \\
\hline
\end{tabular}
\caption{WMAP sources at declination $-60^\circ$ to $-70^\circ$, and with Galactic 
latitude $|b|>10^\circ$.  The five WMAP frequency bands are K (23\,GHz), K$_{\rm a}$ 
(33\,GHz), Q (41\,GHz), V (61\,GHz) and W (94\,GHz), and the flux densities are 
from Bennett et al.\ (2003).  Only one of our sources (J\,2035--6846) was 
detected by WMAP at 94\,GHz, so W--band data are not included in this table. 
$\alpha_{\rm H}$ is the 8--20\,GHz spectral index measured in this paper 
(see \S4), and $\alpha_{\rm WMAP}$ is the WMAP spectral index quoted by Bennett 
et al.\ (2003). 
}
\end{minipage}
\end{table*}

\section{Sources detected by WMAP}
Table 7 lists the sources in our sample which were also detected 
by the WMAP satellite (Bennett et al.\ 2003) at up to five frequencies 
between 23 and 94\,GHz.  All twelve WMAP sources with Galactic latitude 
$|b|>10^\circ$ in the declination zone $-60^\circ$ to $-70^\circ$ were 
detected in our 20\,GHz survey, and Figure 9 shows that there is 
generally good agreement 
between the 23\,GHz flux densities measured by WMAP and the 20\,GHz 
values measured in this study. 
Of the ten sources in our survey with S$_{20}>1$\,Jy, nine are 
also detected by WMAP at 23\,GHz.  The sole exception is 0623--6436, 
a Seyfert 1 galaxy which appears to be strongly variable at 20\,GHz 
(it has a variability index of 43\% in Table 3). 

\begin{figure}
\begin{center}
\includegraphics{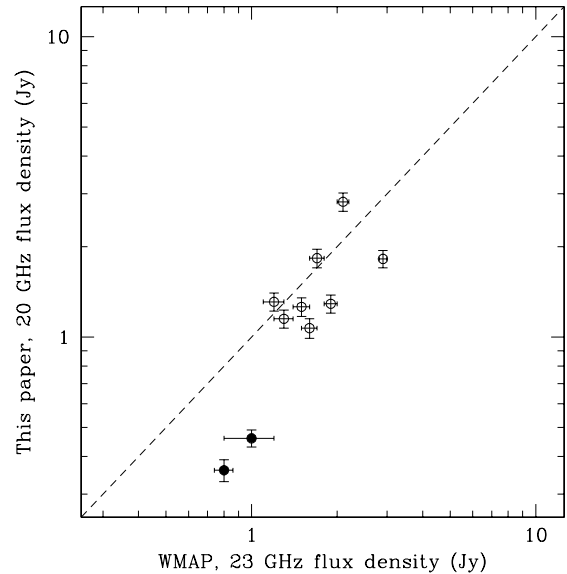} 
\vspace{8cm}
\caption{Comparison of our 20\,GHz flux densities with the 23\,GHz flux 
densities measured by WMAP for sources in common. The two filled circles 
show the objects for which we measured flux--density variability 
over the period 2003--4, as discussed in \S5. }
\end{center}
\end{figure}

\section{Discussion and Conclusions}
Our aim in this paper has been to present some first results on the 
polarization and variability of radio sources selected at 20\,GHz, 
and to outline some methodology which will also be useful in analysing 
data from the much larger ATCA 20\,GHz (AT20G) survey which is now in 
progress. We now summarize and discuss the main results from our pilot 
survey of 173 sources selected to have flux densities above 100\,mJy 
at 20\,GHz. 

\subsection{Radio--source populations at 20\,GHz}

\subsubsection{Comparison with surveys at 15\,GHz}
Our survey is complementary to the recent 15\,GHz studies of Taylor 
et al.\ (2001) and Bolton et al.\ (2004), since we cover a significantly 
larger area of sky, but at a somewhat higher flux--density level. 
It is therefore interesting to compare some of our results with those 
in the 15\,GHz surveys. 
 
In \S5, we used a radio two--colour diagram to characterize the spectra of 
sources selected at 20\,GHz, and showed that most of these sources have 
significant spectral curvature over the range 0.8--20\,GHz.  Similar results 
were found by Bolton et al.\ (2004), who studied the 1.4--43\,GHz radio spectra 
of 176 sources detected at 15\,GHz in a blind survey of about 200\,deg$^2$ 
of sky.  Bolton et al.\ found that 20--30\% of their sources had 
rising spectra at 1--5\,GHz, with the fraction increasing to 39\% for sources 
with S$_{15}>150$\,mJy.  Many of these sources had a spectral peak 
(turnover) above 5\,GHz.  
For our sample selected at 20\,GHz, 39\% of sources with 
S$_{20}>100$\,mJy have rising spectra between 0.8 and 5.0\,GHz 
(i.e. $\alpha_L>0.1$ in Table 5), with the majority of these showing a 
spectral peak above 5\,GHz. Radio sources selected at 20\,GHz 
therefore appear to have similar spectral properties to sources selected  
at the slightly lower frequency of 15\,GHz. 

As noted in \S3.5, the optical identification rate for sources selected at 20\,GHz 
is significantly higher than has been found in surveys to similar flux limits at 
1.4\,GHz.  In contrast to low frequencies, where the strongest radio sources are 
mostly distant, powerful radio galaxies (e.g. Jackson \& Wall 1999), the strongest 
sources at 20\,GHz appear to be mainly QSOs.  A direct comparison 
with the optical results of Bolton et al. (2004) is not completely straightforward, 
since they made most of their optical identifications in the R-band rather than B-band, 
and did not explicitly distinguish between galaxies and candidate QSOs.  However, we 
can directly compare our overall identification rate on the blue SuperCOSMOS 
DSS images with that found by Bolton et al.\ (2004) in the blue Palomar O-band DSS 
images, as shown in Table 8.  Our results confirm the trend found by Bolton et al.\ 
for brighter 15--20\,GHz sources to have a higher optical ID rate and brighter 
optical counterparts. 

\begin{table}
\begin{center}
\begin{tabular}{clccc}
\hline
\multicolumn{1}{c}{Freq.} & \multicolumn{1}{c}{Flux density} & \multicolumn{1}{c}{Median} & 
\multicolumn{2}{c}{Fraction with an optical} \\
\multicolumn{1}{c}{(GHz) } & \multicolumn{1}{c}{limit} & \multicolumn{1}{c}{$b$\,mag.} & 
\multicolumn{2}{c}{ID at $b<21$\,mag} \\
\hline
15 & $>25$\,mJy     & 21.6  &  52$\pm$7\,\% & (64/124)  \\
15 & $>60$\,mJy     & 20.9  &  61$\pm$9\,\% & (43/70)   \\
20 & $>100$\,mJy    & 19.8  &  82$\pm$8\,\% & (118/144) \\
20 & $>500$\,mJy    & 18.9  &  93$\pm$19\,\% & (25/27) \\
\hline
\end{tabular}
\caption{Comparison of the optical properties of radio sources selected in 
high--frequency flux--limited surveys.  The 15\,GHz values are from 
Bolton et al.\ (2004) and the 20\,GHz data from this paper.  }
\end{center}
\end{table}

\subsubsection{Comparison with predictions from low--frequency studies}
Taylor et al.\ (2001) have noted that, as a result of spectral curvature, 
the radio--source population at 15\,GHz cannot be reliably predicted by 
extrapolation from surveys at frequencies of 1--5\,GHz.  Of the sources 
they expected to detect at 15\,GHz, based on extrapolation of the spectral 
index measured from the NVSS 
catalogue at 1.4\,GHz and GB6 catalogue at 5\,GHz, only 45\% (55/122) were 
actually seen.  Furthermore, roughly 10\% of the sources they detected at 
15\,GHz were not predicted by this method. 

We attempted to predict the observed source population in our survey region 
above 100\,mJy at 20\,GHz by extrapolating the radio spectral indices 
measured from the 0.84\,GHz SUMSS 
and 4.85\,GHz PMN surveys.  For a subset of our survey area covering 
about 250 deg$^2$ well away from the Galactic Plane, we detected 33\% (28/84) 
of the sources predicted from extrapolation of the 0.8--5\,GHz spectral 
indices.  Conversely, 18\% (6/34) of the sources we actually detected at 
20\,GHz were not predicted by the low--frequency extrapolation.  
We therefore confirm the findings of Taylor et al.\ (2001) that neither 
the existence nor the flux density of a 15--20\,GHz source can be reliably 
predicted by extrapolating the results of surveys at lower frequencies, 
and show that this is also the case at higher flux densities than were 
probed by the Taylor et al.\ survey. 

\subsubsection{Phase calibrators for ALMA?} 
The results of the previous section are relevant to the planned calibration strategy for 
the ALMA millimetre array now under construction in Chile.  ALMA will operate at 90--720\,GHz, 
and the aim is to calibrate the data by fast--switching between a program source and a 
calibrator less than 1--2 degrees away.  The surface density of currently-known calibrators 
at 90\,GHz is far lower than this, so large numbers of new calibrators will need to 
be found.  The currently-planned strategy for this (Holdaway et al.\ 2004) is to select 
candidates by extrapolation from existing 1--5\,GHz source catalogues.  Our results, 
and those of Taylor et al.\ (2001) imply that such a strategy has at best a 30\% success 
rate in selecting sources at 20\,GHz, and that the success rate at 90\,GHz may be far lower.  

A recent pilot study at 90--100\,GHz of a subset of sources detected in our 20\,GHz 
survey shows that the ATCA can measure accurate continuum flux densities down to levels 
well below 100\,mJy, and that the 8--20\,GHz radio spectral index may be a good predictor 
of the observed flux density at 90\,GHz (Sadler et al., in preparation).  We therefore 
suggest that the 20\,GHz source catalogues now being produced for the whole southern sky
(declination $\delta<0^\circ$)  
in the AT20G survey will provide a more efficient way of identifying 90\,GHz 
phase calibrators for ALMA than the currently--proposed technique of 
extrapolation from radio surveys at 1--5\,GHz. 

\subsubsection{``Flat--spectrum'' and ``steep--spectrum'' populations} 
Earlier in this paper, we characterized the 20\,GHz sources in terms of 
their position in the radio two--colour diagram shown in Figure 7. 
This makes clear both the diversity of radio spectra seen in high--frequency 
sources and the difficulty of predicting high--frequency properties 
from low--frequency spectra.  Studies of the cosmic evolution of radio sources, 
however, usually consider only two source populations --- extended, steep--spectrum 
sources and compact, flat--spectrum sources.  As discussed by Peacock (1985) these 
two populations can be understood physically, with the radio flux density being 
dominated by emission from extended radio lobes in steep--spectrum sources and a 
central compact core in flat--spectrum sources.  
In a sample selected at low frequency, the physically--distinct optically--thin 
diffuse emission has a distribution of spectral indices centred at $-0.7$ and so 
the dividing line between ''flat'' and "steep--spectrum'' sources is traditionally
set at a spectral index of $\alpha=-0.5$.  We note, however, that this division 
will be somewhat frequency--dependent.  

Radio--source samples 
selected at higher frequencies are increasingly dominated by flat--spectrum sources 
which are expected to be compact in nature.  Even at 20\,GHz, however, there is 
a minority population of objects which would be considered steep--spectrum using 
the normal convention of $\alpha<-0.5$ at frequencies of 1--5\,GHz.  
In terms of the overall properties of our sample, and as a guide for later comparison 
with other studies, it is therefore useful to note that our 20\,GHz sample contains 
roughly 87\% ``flat--spectrum'' and 13\% ''steep--spectrum'' as defined by their 
low--frequency spectral index ($\alpha_L$ in Table 5). 

\subsection{Polarization properties at 20\,GHz}
The high selection frequency of our survey makes it particularly useful 
for estimating the contribution of foreground radio sources to future 
studies of polarization fluctuations in the Cosmic Microwave Background 
(CMB) radiation at 20\,GHz and above. 

\begin{figure}
\begin{center}
\includegraphics{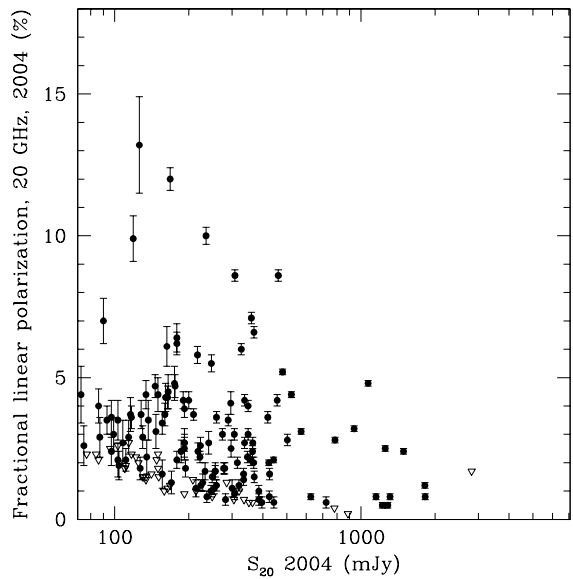} 
\vspace{7.5cm}
\caption{Fractional linear polarization at 20\,GHz, measured in 2004. 
Open triangles show upper limits for sources where no polarized flux 
was detected.  }
\end{center}
\end{figure}

Figure 10 plots the fractional linear polarization measured at 20\,GHz 
against the 20\,GHz flux density, and shows that most sources 
selected at 20\,GHz have low levels of linear polarization 
(typically 1--5\%).  The median fractional polarization at 20\,GHz is 
2.3\%, but Figure 10 suggests that there is a trend for fainter 20\,GHz 
sources to show higher levels of polarization (the median linear polarization 
is 2.7\% for sources with 100\,$<{\rm S}_{20}<$\,200\,mJy and 1.7\% for 
sources with S$_{20}>$\,200\,mJy).  A generalized Kendall's tau correlation test 
for censored data (Isobe et al.\ 1986) gives a value of 2.07, corresponding 
to a 3.9\% probability that the observed correlation is due to chance.   

The median fractional polarization of 2.3\% which we measure at 20\,GHz 
for a flux--limited sample with S$_{20}>100$\,mJy is very close to the 
median value of 2.2\% found by Mesa et al.\ (2002) for a flux--limited 
sample at 1.4\,GHz with S$_{1.4}>80$\,mJy.  Mesa et al.\ also observed 
a marginally--signifcant trend for weaker sources to have a higher 
median polarization. 

The similarity between the median polarizations observed at 1.4\,GHz 
and 20\,GHz is somewhat surprising, since the 1.4\,GHz sample is 
overwhelmingly dominated by steep--spectrum sources and the 20\,GHz 
sample by flat--spectrum sources.  Mesa et al.\ (2002) find a similar 
median polarization at 1.4\,GHz for both steep-- and flat--spectrum 
sources, and Tucci et al.\ (2004) argued that the mean level of polarization 
in flat--spectrum radio sources increases steadily with frequency. 
We might therefore have expected the median polarization in our sample 
to be higher than that observed at 1.4\,GHz.  

\begin{figure}
\begin{center}
\includegraphics{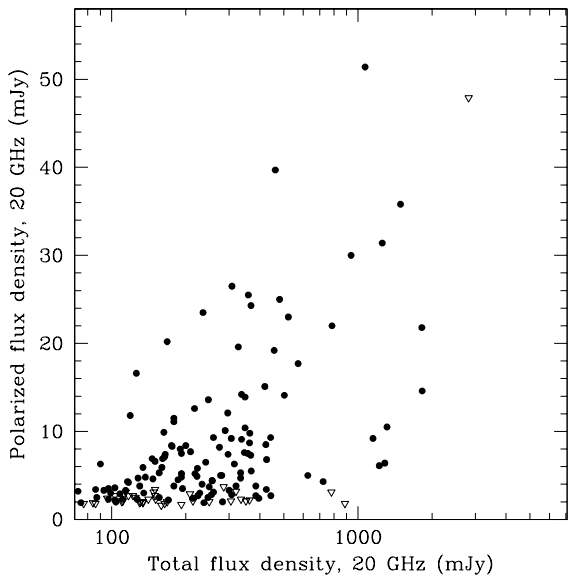} 

\vspace{7.5cm}

\caption{Total linearly--polarized flux density at 20\,GHz, plotted as 
a function of the total flux density. As in Figure 10, open triangles show 
upper limits for sources where no polarized flux was detected.
Note that although the weaker 20\,GHz sources in our sample typically have higher fractional 
polarization, the strongest sources still dominate the source counts 
in polarized flux. 
  }
\end{center}
\end{figure}

This does not appear to be the case, and a larger data set at 20\,GHz 
is needed both to examine this issue in more detail and to compare 
the high--frequency polarization properties of the different spectral 
subclasses identified in \S5 of this paper. 

\subsection{Variability of the source population at 20\,GHz} 
In \S6 of this paper we showed that the general level of variability 
of radio sources selected at 20\,GHz sources appears to be low, 
with a median variability index of 6.9\% on a one--year timescale 
(see Table 6).  In the current sample, we find no significant 
correlation between the variability index of a source and 
its fractional polarization or radio spectral index. 
This is perhaps not surprising, since our sample is relatively small, 
and only a few of the sources are strongly variable.  

The five most variable sources in our sample (with a variability index 
of 30\% or more) are J0507--6104, 
J0623--6436, J0820--6814, J1546--6837 and J1903--6749.  Four of these are 
candidate QSOs of unknown redshift and one (J0623--6436) is a Seyfert galaxy 
at redshift $z$=0.129.  None of these radio sources appear to have been  
monitored previously, so nothing is known about their long-term behaviour. 

Direct comparison of our results with previous studies is difficult, 
both because many of these studies are based on targeted rather than 
flux--limited samples, and because we have so far only analysed data 
from two measurements taken a year apart.  The variability timescales 
measured in this paper are all in the observed frame.  The redshifts 
$z$\ of many of our sources are currently unknown, and we remind the 
reader that the observed variability timescale will differ from the 
intrinsic value by a factor of (1+$z$) so that longer monitoring is 
particularly important for the highest--redshift sources. 

Even at this stage, however, we can conclude that the general level 
of variability in sources selected at 20\,GHz appears to be low 
on timescales of 1--2 years, and that source catalogues made at this 
frequency should therefore be robust on timescales of at least a few 
years.  Long--term monitoring studies of targeted sources by Valtaoja 
and colleagues (e.g. Valtaoja et al.\ 1988) show that even though 
many high--frequency sources have bursts of short--term variability, 
they are relatively quiescent for most of the time.  This is entirely 
consistent with our results, and suggests that we should continue to 
monitor this source sample for a much longer period of time. 

\subsection{Conclusions}
The pilot--study results presented here show that a 
sensitive 20\,GHz radio continuum survey of the whole southern sky 
is feasible, and should produce a uniform source catalogue which is 
largely stable over timescales of a few years.  Such a survey 
should provide further insights into the nature of the 
high--frequency radio--source population, both in its own right 
and as a polarized foreground component in future CMB experiments 
like Planck.

\section{Acknowledgements} 
We acknowledge financial support from the Australian Research Council through 
the award of a Federation Fellowship to RDE and an ARC Australian Professorial 
Fellowship to EMS.  
This research has made use of the NASA/IPAC Extragalactic Database (NED) which 
is operated by the Jet Propulsion Laboratory, California Institute of Technology, 
under contract with the National Aeronautics and Space Administration. 
We thank the referee, Prof.\ Ian Browne, for a number of helpful suggestions.




\appendix

\section[]{Notes on individual sources in Table 3}

{\bf J0025--6028:} \\
Double source with 23\,arcsec separation, PA 159$^\circ$.

\noindent
{\bf J0103--6439:} \\
Wide double at 843\,MHz, with 3.5\,arcmin separation
Only the core is seen at 20\,GHz.  

\noindent
{\bf J0121--6309:} \\
Core plus 39\,arcsec double, PA 14$^\circ$. 

\noindent
{\bf J0257--6112:} \\
Core plus 10\,arcsec jet, PA $\sim$60$^\circ$. 

\noindent
{\bf J0425--6646} \\
This source was identified by Ricci et al. (2004a) with a magnitude 16.8 
stellar object. The higher-resolution radio image we obtined in 2004 makes 
it clear that the correct ID is a fainter stellar object slightly to the west. 

\noindent
{\bf J0715--6829:} \\
This source lies close to a bright (11th magnitude) foreground star, and no optical 
identification is possible from the Supercosmos optical images. 

\noindent
{\bf J0743--6726:} \\
Core plus 12\,arcsec jet, PA 117$^\circ$. 

\noindent
{\bf J1807--7012:} \\
Double source with 27\,arcsec separation, PA 114$^\circ$, 
no core visible. 

\noindent
{\bf J1822--6359:} \\
Double source with 32\,arcsec separation, PA 57$^\circ$. 

\noindent
{\bf J1824--6717:} \\
Double source with 49\,arcsec separation, PA 158$^\circ$, 
no core visible.

\noindent
{\bf J2157--6941:} \\
Core plus wide double source with 1.5\,arcmin separation, PA 20$^\circ$. 
Some flux may be missing at 18 and 20\,GHz. 
This source has been studied in detail by Fosbury et al.\ (1998). 

\noindent
{\bf J2306--6521:} \\ 
Jackson et al.\ (2002) identify this source with a faint (B = 24\,mag.) galaxy, for 
which they measure the quoted redshift of $z$=0.470. 

\noindent
{\bf J2358--6052 and J2350--6057:} \\
Hotspots of the powerful radio galaxy PKS\,2356--61, as discussed by Ricci et al.\ (2004a).



\begin{thebibliography}{99}
\bibitem{} Akritas, M.G., Bershady, M.A., 1996, ApJ, 470, 706 

\bibitem{} Barvainis, R., Lehar, J., Birkinshaw, M., Falcke, H., 
Blundell, K., 2005, ApJ, 618, 122

\bibitem{} Bennett, C.L.\ et al., 2003, ApJ, 583, 1

\bibitem{} Bolton, R.C., Cotter, G., Pooley, G.G., Riley, J.M., Waldram, 
E.M., Chandler, C.J., Mason, B.S., Pearson, T.J., Readhead, A.C.S., 2004, 
MNRAS, 354, 485


\bibitem{} Colless, M.\ et al., 2001, MNRAS, 328, 1039

\bibitem{} Conway, J.E., Cornwell, T.J., Wilkinson, P.N., 1990, 
MNRAS, 246, 490 

\bibitem{} De Zotti, G., Ricci, R., Mesa, D., Silva, L., Mazzotta, P.,
Toffolatti, L., Gonzalez--Nuevo, J., 2005, A\&A. 431, 893

\bibitem{} Fosbury, R.A.E.,  Morganti, R.,  Wilson, W., Ekers, R.D., 
di Serego Alighieri, S., Tadhunter, C.N., 1998, MNRAS 296, 701. 

\bibitem{} Hambly, N.C.\ et al., 2001, MNRAS, 326, 1279

\bibitem{} Harris, A.I., Zmuidzinas, J., 2001, Rev. Sci. Instrum. 72, 1531 

\bibitem{} Hirabayashi, H. et al., 2000, PASJ 52, 997

\bibitem{} Holdaway, M., Carilli, C., Laing, R., 2004, 
ALMA Memo 493: Finding Fast Switching Calibrators for ALMA, 
http://www.alma.nrao.edu/memos/

\bibitem{} Isobe, T., Feigelson, E.D., Nelson, P.I., 1986, ApJ, 306, 490

\bibitem{} Jackson, C.A., Wall, J.V., 1999, in ``Looking Deep in the Southern Sky'', 
ed. R.\ Morganti \& W.J.\ Couch. Berlin: Springer-Verlag, p. 11

\bibitem{} Jackson, C.A., et al., 2002, A\&A, 386, 97

\bibitem{} Leahy, P., 1989, VLA Scientific Memorandum No.\ 161, 
NRAO 

\bibitem{} Lo, K.Y. et al.\ 2001, in 
{\it 20th Texas Symposium on Relativistic Astrophysics}, 
eds., J. Craig Wheeler and Hugo Martel, AIP Conf. Proc., 586, 172  

\bibitem{} Ma, C.\ et al.\ 1998, AJ, 116, 516

\bibitem{} Mauch, T., Murphy, T., Buttery, H.J., Curran, J., Hunstead, R.W., 
Piestrzynski, B., Robertson, J.G., Sadler, E.M., 2003. MNRAS, 342, 1117

\bibitem{} Mesa, D., Baccigalupi, C., De Zotti, G., Gregorini, L., Mack,
K.-H., Vigotti, M., Klein, U., 2002, A\&A, 396, 463

\bibitem{} Owen, F.N., Spangler, S.R., Cotton, W.D., 1980, 
AJ, 85, 3510 

\bibitem{} Peacock, J.A, Gull, S.F., 1981, MNRAS, 194, 331 

\bibitem{} Peacock, J.A., 1985, MNRAS, 217, 601 

\bibitem{} Reynolds, J.E., 1994, ATNF Memo AT/39.3/040,  
http://www.\\*atnf.csiro.au/observers/memos/d96783\~1.pdf

\bibitem{} Ricci, R., Sadler, E.M., Ekers, R.D., Staveley--Smith, L., 
 Wilson, W.E., Kesteven, M.J., Subrahmanyan, R., Walker, M.A., Jackson, 
C.A., De Zotti, G., 2004a, MNRAS, 354, 305 

\bibitem{} Ricci, R., Prandoni, I., Gruppioni, C., Sault, R.J., 
De Zotti, G., 2004b, A\&A, 415, 549 

\bibitem{} Roberts, P.P., Leach, M.R., Wilson, W.E., 2006, in preparation 

\bibitem{} Sadler, E.M.\ et al.\ 2002, MNRAS, 329, 227

\bibitem{} Sault, R.J., Wieringa, M.H., 1994, A\&AS, 108, 585 

\bibitem{} Sault, R.J., 2003, unpublished ATNF Memo,  
http://www.\\*narrabri.atnf.csiro.au/calibrators/data/1934-638/\\*1934\_12mm.pdf

\bibitem{} Taylor, A.C., Grainge, K., Jones, M.E., Pooley, G.G., Saunders, 
R.D.E., Waldram, E.M., 2001, MNRAS, 327, L1

\bibitem{} Tingay, S.J., Jauncey, D.L., King, E.A., Tzioumis, A.K., Lovell, J.E.J., 
Edwards, P.G., 2003, PASJ, 55, 351

\bibitem{} Tornikoski, M., Lahteenmaki, A., Lainela, M., Valtaoja, E., 2002, 
ApJ, 579, 136 

\bibitem{} Tucci, M., Martínez-Gonzalez, E., Toffolatti, L., Gonzalez-Nuevo, J., 
De Zotti, G., 2004, MNRAS, 349, 1267 

\bibitem{} Valtaoja, E.\ et al., 1988, A\&A, 203, 1

\end{thebibliography}
\end{document}